\documentclass[aps,prd,superscriptaddress,amsmath,preprint]{revtex4}

\usepackage{enumerate}
\usepackage{amssymb}
\usepackage{amsfonts}
\usepackage{latexsym}
\usepackage{bm}
\usepackage{amscd}
\usepackage{graphicx}
\usepackage{epsfig}
\usepackage{hhline,multirow}
\usepackage{dcolumn}
\usepackage{url}
\usepackage[colorlinks=true,linkcolor=blue]{hyperref}
\usepackage{color}
\usepackage{indentfirst}
\usepackage{subfigure}
\usepackage{psfrag}
\usepackage{slashed}
\usepackage{float}
\usepackage{setspace}

\preprint{ADP-16-35/T991}
\preprint{JLAB-THY-16-2357}

\begin{document}

\title{Strange quark asymmetry in the proton in chiral effective theory}

\author{X.~G. Wang}
\affiliation{CoEPP and CSSM, University of Adelaide,
                Adelaide 5005, Australia}
\author{Chueng-Ryong Ji}
\affiliation{North Carolina State University,
                Raleigh, North Carolina 27695, USA}
\author{W. Melnitchouk}
\affiliation{Jefferson Lab, Newport News,
                Virginia 23606, USA}
\author{Y. Salamu}
\affiliation{Institute of High Energy Physics, CAS,
                Beijing 100049, China}
\author{A. W. Thomas}
\affiliation{CoEPP and CSSM, University of Adelaide,
                Adelaide 5005, Australia}
\author{P. Wang}  
\affiliation{Institute of High Energy Physics, CAS,
                Beijing 100049, China}
\affiliation{Theoretical Physics Center for Science Facilities, CAS,
                Beijing 100049, China}

\date{\today}

\begin{abstract}
We perform a comprehensive analysis of the strange--antistrange
parton distribution function (PDF) asymmetry in the proton in the
framework of chiral effective theory, including the full set of
lowest order kaon loop diagrams with off-shell and contact
interactions, in addition to the usual on-shell contributions
previously discussed in the literature.
We identify the presence of $\delta$-function contributions to
the $\bar s$ PDF at $x=0$, with a corresponding valence-like
component of the $s$-quark PDF at larger $x$, which allows
greater flexibility for the shape of $s-\bar s$.
Expanding the moments of the PDFs in terms of the pseudoscalar
kaon mass, we compute the leading nonanalytic behavior of the
number and momentum integrals of the $s$ and $\bar s$ distributions,
consistent with the chiral symmetry of QCD.
We discuss the implications of our results for the understanding of
the NuTeV anomaly and for the phenomenology of strange quark PDFs
in global QCD analysis.
\end{abstract}

\maketitle

\section{Introduction}
\label{sec.intro}

Historically, the simplest quark models envisaged the nucleon's
properties and structure being determined entirely in terms of its
valence $u$- and $d$-quark constituents.  The subsequent development
of QCD necessitated refinements of this picture, in which a sea of
virtual quark--antiquark ($q\bar q$) pairs and gluons made the nucleon
a far richer and more dynamic environment.  In this new paradigm,
not only did the light-quark $q\bar q$ sea display nontrivial
structure, but heavier quarks such as the strange or even charm
quark could contribute locally to the internal nucleon dynamics.

The role that strange quarks, in particular, play in the nucleon has
been the focus of attention in hadronic physics for nearly three
decades.  Early polarized deep-inelastic scattering (DIS) experiments
suggested that a surprisingly large fraction of the proton's spin
might be carried by strange quarks \cite{EMC89}, in contrast to the
naive quark model expectations \cite{EJ74}.
Recognition that the spatial distributions of strange quarks and
antiquarks could be different further motivated searches for strange
contributions to the nucleon's electroweak form factors
\cite{Ahrens87, Garvey93, Kaplan88, McKeown89, Beck89, Musolf94}.
Dedicated programs of strange form factor measurements through
parity-violating electron scattering at Jefferson Lab and other
facilities \cite{Beise05, Paschke11, Armstrong12} subsequently
yielded very precise determinations of both the strange electric
and magnetic form factors of the nucleon \cite{Young06}, enabling
rigorous comparisons with lattice QCD and chiral effective theory
\cite{AdlGEs, AdlGMs}, as well as fundamental tests of the
Standard Model \cite{Young07}.

One of the guiding principles for understanding the nonperturbative
features of strange quarks and antiquarks in the nucleon sea has been
chiral symmetry breaking in QCD.  While the generation of $s \bar s$
pairs through perturbative gluon radiation typically produces symmetric
$s$ and $\bar s$ distributions (at least up to two loop corrections
\cite{Catani04}), any significant difference between the momentum
dependence of the $s$ and $\bar s$ parton distribution functions
(PDFs) would be a clear signal of nonperturbative effects.
In fact, insights from chiral symmetry breaking in the nonstrange
sector led to the prediction \cite{Thomas83} of an excess of
$\bar d$ antiquarks over $\bar u$ in the proton, which was
spectacularly confirmed in DIS \cite{NMC, HERMES} and Drell-Yan
\cite{NA51, E866} experiments more than a decade later.
A similar mechanism, which can be intuitively realized in the
form of a pseudoscalar meson cloud surrounding a valence-quark
nucleon core, was subsequently used \cite{Signal87} to demonstrate
the natural emergence of a nonzero $s-\bar s$ asymmetry from the
breaking of the chiral SU(3) symmetry of QCD.

While the existence of an $s-\bar s$ asymmetry is not, from the
point of view of nonperturbative QCD dynamics, terribly surprising
in itself, the magnitude and even the sign of the asymmetry has
historically been far more difficult to determine.
Experimentally, from an analysis of $\nu$ and $\bar\nu$ DIS data
from the BEBC, CDHS and CDHSW experiments, Barone {\it et al.}
\cite{Zomer00} concluded that the $s$-quark PDF was somewhat
harder than the $\bar s$.  Quantitatively, the second moment of
the asymmetry,
\begin{eqnarray}
S^- &=& \int_0^1 dx\, x\, \big( s(x)-\bar s(x) \big),
\label{eq:S-}
\end{eqnarray}
where $x$ is the light-cone momentum fraction of the nucleon
carried by the strange parton, was constrained to be
	$S^- \approx (2 \pm 3) \times 10^{-3}$.
Of course, by strangeness conservation the first moment of
$s-\bar s$ must vanish identically, which, in the absence of
contributions at $x=0$, would suggest the presence of at least
one zero in the $x$ dependence of $s-\bar s$ at finite $x$.
Analysis of more recent CCFR \cite{CCFR} and NuTeV \cite{Zeller02}
data on opposite sign dimuon production in neutrino--nucleus DIS
yielded \cite{Zeller02} a negative asymmetry,
        $S^- = (-2.7 \pm 1.3) \times 10^{-3}$,
at leading order, although a later, next-to-leading order analysis
\cite{Mason07} found positive values,
        $S^- = (1.96 \pm 1.43) \times 10^{-3}$
at $Q^2 = 16$~GeV$^2$.

Beyond extractions from individual experiments, global QCD analyses
of charged lepton and neutrino DIS, along with other high energy
scattering data, have generally found positive values for $S^-$.
On the other hand, the various approximations made about nuclear
corrections to the neutrino data and the various functional forms
chosen for the PDFs make any current phenomenological analysis
subject to sizeable uncertainties.
Taking into account some of these uncertainties, the phenomenological
analysis of Bentz {\it et al.} \cite{Bentz10} concluded that
        $S^- = (0 \pm 2) \times 10^{-3}$ at $Q^2=16$~GeV$^2$.

While the current empirical situation with $S^-$ remains somewhat
inconclusive, a number of theoretical estimates have been made,
based on perturbative and nonperturbative QCD arguments.
Catani {\it et al.} \cite{Catani04}, for instance, showed that
perturbative three-loop effects can induce nonzero negative $S^-$
values,
        $S^- \approx -0.5 \times 10^{-3}$,
through $Q^2$ evolution of symmetric $s/\bar s$ distributions
from a low input scale, $Q_0 \approx 0.5$~GeV.  
Nonperturbatively, the most common approach to computing the
$s-\bar s$ asymmetry has been in the framework of meson cloud
models, which focus on the role of the nucleon's light-front
wave function with Fock state component consisting of kaons and
hyperons, $Y = \Lambda, \Sigma, \ldots$  Here the asymmetric
dissociation of the nucleon into a hyperon (containing the $s$
quark) and a kaon (containing the $\bar s$ antiquark) automatically
generates asymmetric distributions for the $s$ and $\bar s$ PDFs.

First estimated nearly 3 decades ago using phenomenological    
nucleon-kaon-hyperon vertex form factors \cite{Signal87},      
subsequent kaon cloud model calculations have, however, at times
yielded conflicting results.
Using a light-front formalism that enabled simultaneous computation
of strange observables in both deep-inelastic and elastic scattering,
the small experimental values of the strange electromagnetic form
factors were found \cite{Malheiro97, Malheiro99} to restrict the
magnitude of $s-\bar s$ to be very small, with a shape strongly
dependent on the choice of the $NKY$ vertex function.
Cao and Signal \cite{Cao03} later observed that while fluctuations
to $K \Lambda$ and $K \Sigma$ states gave rise to a small positive
asymmetry,
        $S^- = 0.143 \times 10^{-3}$,
the inclusion of the heavier $K^*$ mesons \cite{Barz98} changed the  
sign of the overall asymmetry,
        $S^- = -0.135 \times 10^{-3}$,    
with the magnitude remaining rather small.
Considering $K \Lambda$ fluctuations of the nucleon with a Gaussian
probability distribution whose parameters are constrained by      
inclusive DIS data and normalization tuned to $x(s+\bar s)$ from
the CCFR data \cite{CCFR}, Alwall and Ingelman \cite{Alwall04}     
found a harder $s$ PDF than $\bar s$, with
        $S^- = 1.65 \times 10^{-3}$.
In that model the fluctuations to $K \Sigma$ and $K^* \Lambda$ 
states were argued to be implicitly included in the $K \Lambda$  
result, with the sign of $S^-$ remaining positive.

Models with couplings to the mesons parametrized at the quark
level have also been considered by several authors.
Using an effective chiral quark model with constituent quarks
coupling to Goldstone bosons, Ding {\it et al.} \cite{Ma05}
found
        $S^- \approx (4-9) \times 10^{-3}$,
depending on the input used for bare constituent quark         
distributions.
Wakamatsu \cite{Wakamatsu05} used an SU(3) chiral quark soliton    
model with an effective mass difference parameter between the
strange and nonstrange quarks to obtain the range
        $S^- = (2.5-5.5) \times 10^{-3}$.
Most recently, Hobbs {\it et al.} extended previous light-front
calculations using a scalar tetraquark spectator model with
Gaussian and power-law wave functions \cite{Hobbs15}, finding     
        $S^- = (-1~\rm to~+5) \times 10^{-3}$.

In all these models, while the basic physics principles underlying
the generation of the $s-\bar s$ asymmetry are similar, the
{\it ad hoc} nature of some of the model assumptions and ingredients
have inevitably led to a fairly wide range of predictions, with a
consequent lack of consensus about the nature of the asymmetry.
Clearly, if one is to make reliable predictions for $S^-$, a more  
systematic approach is needed, one which has a more direct        
connection to the underlying QCD theory.

The first such unambiguous connection between the kaon cloud of the
nucleon and QCD came with the realization \cite{TMS00} that in chiral
expansions of moments of strange quark PDFs, the coefficients of the
leading nonanalytic (LNA) terms in the kaon mass, $m_K$, are model
independent and can only arise from pseudoscalar meson loops.
Starting from the most general effective Lagrangian consistent with
the chiral symmetry of QCD, at a given order in the chiral expansion a
unique set of diagrams can be identified and computed systematically
\cite{Arndt01, Chen02, Detmold01}.
The long distance ($m_K \to 0$) effects in such expansions are
thus dictated solely by chiral symmetry and gauge invariance,
while the short distance contributions are treated with a
particular regularization procedure.  The choice of regularization
scheme introduces additional parameters into the calculation,
which can be fixed by comparing with specific observables.

This methodology was applied by Salamu {\it et al.} \cite{Salamu15}
to the case of pion loops and their effects on the $\bar d-\bar u$
asymmetry in the proton, using for illustration a simple sharp cutoff
on the transverse momentum of the pion $k_\perp$ for the ultraviolet
regulator.
More recently Wang {\it et al.} \cite{XWang16} generalized this
approach to the SU(3) sector, using Pauli-Villars (PV) regularization
to compute the various lowest order diagrams in the chiral SU(3)
expansion, and obtain a range for $S^-$ consistent with available
phenomenological constraints.

In the present work, we extend the analysis of Ref.~\cite{XWang16},
providing full details of the calculation of the kaon loop
contributions to the strange-quark PDF and its moments in the
chiral effective theory.  We outline the formal derivation of
the convolution representation, and perform a numerical study
of the various contributions from the lowest order diagrams.
We emphasize the importance of using regularization procedures
that preserve the chiral and gauge symmetries of QCD, and
contrast these with previous calculations in the literature
using form factors at hadronic vertices.

We further explore the consequences of the $\delta$-function
contribution to the $\bar s$ distribution at zero momentum fraction that
arises from the Weinberg-Tomozawa contact interaction in the chiral
theory, and identify a valence-like component of the strange PDF.
Suggestions of possible $\delta$-function contributions to PDFs were
raised earlier \cite{Broadhurst73, Bass05} in discussions of the
unpolarized Schwinger term and proton spin sum rules.
The practical implication of the $\delta$-function terms is
to provide significantly greater flexibility in the allowed
phenomenological parametrization of the $s-\bar s$ difference,
suggesting that current forms used in global PDF analysis may be
too restrictive.

In addition to its intrinsic value, understanding the sign and
magnitude of the $s-\bar s$ asymmetry is also vital for the extraction
of the Paschos-Wolfenstein ratio from neutrino--nucleus DIS data.
Specifically, it has been suggested that a large positive value of
$S^- \sim 2 \times 10^{-3}$ could resolve much of the discrepancy
between the $\sin^2\theta_W$ value extracted by the NuTeV
Collaboration \cite{NuTeV} and the Standard Model \cite{Bentz10}.
A negative value for $S^-$ would, in contrast, exacerbate the
disagreement.  Thus, an accurate determination of the magnitude,
as well as the sign, of $S^-$ would be of significant practical
value in resolving this issue.

In Sec.~\ref{sec:L} we begin by defining the chiral SU(3) Lagrangian,
identifying the terms at the lowest order in the expansion that
contribute to the strange quark distributions in the nucleon.
The details of the computation of nucleon PDFs and their moments
within the effective chiral theory framework are presented in
Sec.~\ref{sec:formalism}.  Here we discuss the matching of the
quark-level operators with the corresponding hadronic operators,
the coefficients of which are related to moments of specific PDFs.
The operator formalism is also shown to lead to a natural represention
of the nucleon PDFs in the form of convolutions of PDFs of hadronic
constituents and nucleon $\to$ hadron splitting functions (or hadronic
light-cone momentum distributions).
Explicit expressions for the latter are derived in Sec.~\ref{sec:fy}
for each kind of kaon and hyperon splitting function allowed at the
lowest order, including the kaon and hyperon rainbow, kaon bubble and
tadpole, and Kroll-Ruderman vertex contributions.
In Sec.~\ref{sec:model-indep} the model-independent features of the
kaon loop corrections to the $s$ and $\bar s$ PDFs are discussed.
Expanding the moments of the PDFs in powers of the kaon mass, we
identify the leading nonanalytic behavior of the lowest two moments,
which is a unique and model-independent feature of pseudoscalar loops
that all calculations consistent with QCD must respect.

The regularization of the hadronic splitting functions is discussed
in Sec.~\ref{sec:reg}.  We review the PV prescription,
which was shown in Ref.~\cite{XWang16} to be a viable method,
consistent with chiral and gauge symmetry, for obtaining consistent
results in terms of a small number of cutoff parameters fixed from
phenomenology.  In addition, we explore other regularization schemes,
such as using phenomenological vertex form factors.  While naive
application of hadronic form factors leads to problems with gauge
invariance, we illustrate a nonlocal approach which allows the
symmetry to be preserved.
The numerical results for the strange and antistrange PDFs in
the nucleon are presented in Sec.~\ref{sec:results}.
The magnitude and sign of the strange--antistrange asymmetry
are determined by cutoff parameters that are constrained by
other observables, such as hyperon production in inclusive $pp$
scattering, that are sensitive to the presence of strangeness
in the nucleon, as well as information from global PDF analyses.
Using all available constraints from data, we obtain upper and
lower limits on the second moment of $s-\bar s$, and discuss its
impact on the NuTeV anomaly.
Finally, in Sec.~\ref{sec:conc} we summarize our findings and
outline possible future improvements in theory and experiment
that can lead to a better understanding of the
strange asymmetry in the nucleon.

\section{Chiral effective Lagrangian}
\label{sec:L}

The effective Lagrangian for the interaction of octet baryons $B$
through pseudoscalar fields $\phi$, consistent with chiral SU(3)
symmetry, can be written at lowest order in the derivative expansion
as \cite{Jenkins91, Bernard08, Shanahan13}
\begin{eqnarray}\label{eq:chiral-lagrangian}
\mathcal{L} &=&
 - \frac{D}{2} \bar{B} \gamma_\mu \gamma_5\, \{u^\mu,B\}
 - \frac{F}{2} \bar{B} \gamma_\mu \gamma_5\, [u^\mu,B]
 + i\, \bar{B} \gamma_{\mu}\, [D^{\mu},B],
\end{eqnarray}
where 
\begin{equation}
u_\mu = i \left( u^\dag\, \partial_\mu u - u\, \partial_\mu u^\dag
	  \right),
\end{equation}
and the operator $u$ is given in terms of the pseudoscalar fields by
\begin{equation}
u = \exp \left( \frac{i\phi}{\sqrt{2} f_\phi} \right),
\end{equation}
with $f_\phi$ the pseudoscalar decay constant.
The covariant derivative $D^\mu$ is defined as
\begin{equation}
[D^\mu, B] = \partial_\mu B + [\Gamma_\mu, B],
\end{equation}
and $\Gamma_\mu$ is the link operator,
\begin{equation}
\Gamma_\mu = \frac{1}{2} [u^\dag,\partial_\mu u].
\end{equation}
The constants $D$ and $F$ in Eq.~(\ref{eq:chiral-lagrangian}) are
the SU(3) flavor coefficients associated with the anticommutator
and commutator of $u^\mu$ and $B$, respectively.

The pseudoscalar field $\phi$ can be written explicitly in
matrix form in terms of the isovector $\pi$, isodoublet $K$,
and isosinglet $\eta$ fields as
\begin{eqnarray}
\phi\ =\ \sum_{a=1}^8 \frac{\lambda_a}{\sqrt{2}} \phi_a
&=& \left(
    \begin{array}{ccc}
      \frac{1}{\sqrt{2}}\pi^0 + \frac{1}{\sqrt{6}}\eta
	& \pi^+
	& K^+ 				\\
      \pi^-
	& -\frac{1}{\sqrt{2}}\pi^0 + \frac{1}{\sqrt{6}}\eta
	& K^0				\\
      K^-
	& \overline{K}^0
	& -\frac{2}{\sqrt{6}}\eta	\\
    \end{array}
    \right),
\label{eq:phi}
\end{eqnarray}  
where $\lambda_a$ are the SU(3) Gell-Mann matrices, and the
fields $\phi_a$ are given by
\mbox{$\phi_1 = (\pi^+ + \pi^-)/\sqrt{2}$},
$\phi_2 = i (\pi^+ - \pi^-)/\sqrt{2}$,
$\phi_3 = \pi^0$,
$\phi_4 = (K^+ + K^-)/\sqrt{2}$,
\mbox{$\phi_5 = i (K^+ - K^-)/\sqrt{2}$},
$\phi_6 = (K^0 + \overline{K}^0)/\sqrt{2}$,
$\phi_7 = i (K^0 - \overline{K}^0)/\sqrt{2}$, and
$\phi_8 = \eta$.
Similarly, the octet baryon field $B$ can be expressed in terms of
the nucleon, the strangeness $-1$ hyperons $\Sigma$ and $\Lambda$,
and the strangeness $-2$ hyperon $\Xi$ fields as
\begin{eqnarray}
B\ =\ \sum_{a=1}^8 \frac{\lambda_a}{\sqrt{2}} B_a
&=& \left(
    \begin{array}{ccc}
      \frac{1}{\sqrt{2}}\Sigma^0 + \frac{1}{\sqrt{6}}\Lambda
	& \Sigma^+
	& p				\\
      \Sigma^-
	& -\frac{1}{\sqrt{2}}\Sigma^0+\frac{1}{\sqrt{6}}\Lambda
	& n				\\
      \Xi^-
	& \Xi^0
	& -\frac{2}{\sqrt{6}}\Lambda	\\
    \end{array}
    \right),
\label{eq:B}
\end{eqnarray}
where the assignment of the individual baryon fields $B_a$ is
$B_1 = (\Sigma^+ + \Sigma^-)/\sqrt{2}$,
\mbox{$B_2 = i (\Sigma^+ - \Sigma^-)/\sqrt{2}$},
$B_3 = \Sigma^0$,
$B_4 = (p + \Xi^-)/\sqrt{2}$,
$B_5 = i (p - \Xi^-)/\sqrt{2}$,
$B_6 = (n + \Xi^0)/\sqrt{2}$,
$B_7 = i (n - \Xi^0)/\sqrt{2}$, and
$B_8 = \Lambda$.

For practical applications, in the following we will restrict ourselves
to the case of a nucleon initial state, although the generalization to
hyperon initial states is straightforward.
Using the representations (\ref{eq:phi}) and (\ref{eq:B}), the chiral
Lagrangian ${\cal L}$ in Eq.~(\ref{eq:chiral-lagrangian}) can be
expanded to ${\cal O}((\phi/f_\pi)^2)$ as a sum of terms involving
a single pseudoscalar meson coupling to the baryon current,
	${\cal L}_{\phi BB}$,
and a Weinberg-Tomozawa term,
	${\cal L}_{\phi\phi BB}$,
in which two pseudoscalar mesons couple to the baryon at the same point,
	${\cal L} = {\cal L}_{\phi BB} + {\cal L}_{\phi\phi BB}$.
The former generates the well-known ``rainbow'' diagram, in which
a pseudoscalar meson is emitted and reabsorbed by the baryon at
different space-time points,
\begin{eqnarray}
\mathcal{L}_{\phi BB}
&=& \frac{1}{2 f_\phi}
\Big\{ (D+F)
\Big[
   \bar{p} \gamma^\mu \gamma_5 p\, \partial_\mu \pi^0
 - \bar{n} \gamma^\mu \gamma_5 n\, \partial_\mu \pi^0
 + \sqrt{2}
   (\bar{n} \gamma^\mu \gamma_5 p\, \partial_\mu \pi^-
   +\bar{p} \gamma^\mu \gamma_5 n\, \partial_\mu \pi^+)
\Big]					\nonumber\\
&+& (D-F)
\Big[
   \overline{\Sigma}^0 \gamma^\mu \gamma_5 p\, \partial_\mu K^-
 + \bar{p} \gamma^\mu \gamma_5 \Sigma^0\, \partial_\mu K^+
 + \sqrt{2}
   (\overline{\Sigma}^+ \gamma^\mu \gamma_5 p\, \partial_\mu \overline{K}^0
   +\bar{p} \gamma^\mu \gamma_5 \Sigma^+\, \partial_\mu K^0)
\Big]					\nonumber\\
&-& (D-F)
\Big[
   \overline{\Sigma}^0 \gamma^\mu \gamma_5 n\, \partial_\mu \bar{K}^0
 + \bar{n} \gamma^\mu \gamma_5 \Sigma^0\, \partial_\mu K^0
 - \sqrt{2}
   (\overline{\Sigma}^- \gamma^\mu \gamma_5 n\, \partial_\mu K^-
   +\bar{n} \gamma^\mu \gamma_5 \Sigma^-\, \partial_\mu K^+)
\Big]					\nonumber\\
&-& \frac{1}{\sqrt{3}} (D+3F)
\Big[
   \overline{\Lambda} \gamma^\mu \gamma_5 p\, \partial_\mu K^-
  +\bar{p} \gamma^\mu \gamma_5 \Lambda\, \partial_\mu K^+
  + \overline{\Lambda} \gamma^\mu \gamma_5 n\, \partial_\mu \bar{K}^0
  +\bar{n} \gamma^\mu \gamma_5 \Lambda\, \partial_\mu K^0
\Big]                                  \nonumber\\
&-& \frac{1}{\sqrt{3}} (D-3F)
\Big[
   \bar{p} \gamma^\mu \gamma_5 p\, \partial_\mu \eta
 + \bar{n} \gamma^\mu \gamma_5 n\, \partial_\mu \eta 
 \Big]
\Big\}.
\label{eq:LpBB}
\end{eqnarray}
The latter term,
\begin{eqnarray}
\mathcal{L}_{\phi\phi BB} &=& \frac{i}{(2 f_\phi)^2}	\nonumber\\
&\times&
\Big\{
  \bar{p} \gamma^\mu p
  \left[ \pi^+ \partial_\mu \pi^- - \pi^- \partial_\mu \pi^+
       + 2 (K^+ \partial_\mu K^- - K^- \partial_\mu K^+)
       + K^0 \partial_\mu \overline{K}^0
	 - \overline{K}^0 \partial_\mu K^0
  \right]						\nonumber\\
& & \hspace{-0.2cm} +\,
  \bar{n} \gamma^\mu n
  \left[ \pi^- \partial_\mu \pi^+ - \pi^+\partial_\mu \pi^-
       + K^+ \partial_\mu K^- - K^- \partial_\mu K^+
       + 2 (K^0 \partial_\mu \overline{K}^0
	 - \overline{K}^0 \partial_\mu K^0)
  \right]   \nonumber\\
& &  \hspace{-0.2cm} 
+\, 
   \bar{p} \gamma^\mu n
  \left[ \sqrt{2}(\pi^0 \partial_\mu \pi^+ - \pi^+\partial_\mu \pi^0)
       + K^+ \partial_\mu \overline{K}^0 - \overline{K}^0 \partial_\mu K^+
  \right]   \nonumber\\
  & & \hspace{-0.2cm} 
+\,
  \bar{n} \gamma^\mu p
  \left[ \sqrt{2}(\pi^- \partial_\mu \pi^0 - \pi^0\partial_\mu \pi^-)
       + K^0 \partial_\mu K^- - K^- \partial_\mu K^0
  \right]
  \Big\},
\label{eq:LppBB}
\end{eqnarray}
is necessary for the preservation of chiral symmetry, and is
independent of the couplings $D$ and $F$.
The effective interactions in Eqs.~(\ref{eq:LpBB}) and (\ref{eq:LppBB})
then form the basis for the derivation of the effective hadronic
operators, whose matrix elements will be related to moments of PDFs.

\section{PDFs in chiral effective theory}
\label{sec:formalism}

From the effective chiral Lagrangian we can derive expressions for
parton distributions in the nucleon by matching twist-two quark
operators with the hadronic operators in the effective theory.
The matrix elements of these operators are then related through
the operator product expansion in QCD to moments of the PDFs.
In this section we present the formalism needed for the analysis
of the PDF moments and identify the complete set of hadronic
operators relevant for the computation of the strange-quark
distribution in the nucleon.

\subsection{Convolution formalism}
\label{ssec:conv}

We begin by defining the $n$-th Mellin moment ($n \geq 1$) of a
spin-averaged PDF $q(x)$ in the nucleon for a given flavor $q$
($q = u, d, s, \ldots$) by
\begin{eqnarray}
\langle x^{n-1} \rangle_q
&=& \int_{-1}^1 dx\, x^{n-1}\, q(x)		\nonumber\\
&=& \int_0^1 dx\, x^{n-1}
    \Big( q(x) + (-1)^n\, \bar{q}(x) \Big),
\label{eq:mom_def}
\end{eqnarray}
where the sign on the antiquark contribution $\bar q(x)$ reflects
the crossing symmetry properties of the spin-averaged PDFs,
$q(-x) = -\bar q(x)$, and for brevity we suppress explicit
dependence of the PDFs on the scale $Q^2$.
The operator product expansion allows these moments to be
related to the matrix elements of local twist-two operators
$\mathcal{O}_q^{\mu_1 \cdots \mu_n}$ between nucleon states
with momentum~$p$,
\begin{equation}
\langle N(p) | \mathcal{O}_q^{\mu_1 \cdots \mu_n} | N(p) \rangle
= 2\, \langle x^{n-1} \rangle_q\, 
  p^{\mu_1} \cdots p^{\mu_n},
\end{equation}
where the spin-$n$ operators are given by quark bilinears
\begin{equation}\label{eq:Oq}
\mathcal{O}^{\mu_1 \cdots \mu_n}_q
= i^{n-1}\, \bar{q} \gamma^{ \{ \mu_1 }\overleftrightarrow{D}^{\mu_2}
  \cdots \overleftrightarrow{D}^{ \mu_n \} } q\, ,
\end{equation}
with $\overleftrightarrow{D}
= \frac{1}{2} \big( \overrightarrow{D} - \overleftarrow{D} \big)$,
and the braces $\{\, \cdots \}$ indicate total symmetrization of
Lorentz indices.

In the effective field theory, the quark operators ${\cal O}_q$ are
matched to hadronic operators ${\cal O}_j$ having the same quantum
numbers (but not necessarily the same twist) \cite{Chen02},
\begin{equation}
\mathcal{O}^{\mu_1 \cdots \mu_n}_q
= \sum_{j} c^{(n)}_{q/j}\ \mathcal{O}^{\mu_1 \cdots \mu_n}_j,
\label{eq:match}
\end{equation}
where $j$ labels different types of hadronic operators, and the
coefficients $c^{(n)}_{q/j}$ are the $n$-th moments of the PDF
$q_j(x)$ in the hadronic configuration $j$,
\begin{eqnarray}
c^{(n)}_{q/j}
&=& \int_{-1}^1 dx\, x^{n-1}\, q_j(x)\,
\equiv\, \langle x^{n-1} \rangle_{q/j}.
\end{eqnarray}
The nucleon matrix elements of the hadronic operators
${\cal O}_j^{\mu_1 \cdots \mu_n}$ can be written in terms of
moments of the hadronic $N \to j$ splitting functions $f_j(y)$,
\begin{equation}
\langle N(p) | \mathcal{O}_j^{\mu_1 \cdots \mu_n} | N(p) \rangle
= 2\, f_j^{(n)}\,
  p^{ \{ \mu_1 } \cdots p^{ \mu_n\} },
\label{eq:fjn_def}
\end{equation}
where the moment $f_j^{(n)}$ is given by the integral
\begin{eqnarray}
f_j^{(n)}
&=& \int_{-1}^{1} dy\, y^{n-1} f_j(y),
\label{eq:fjn}
\end{eqnarray}
with $y$ the light-cone momentum fraction of the nucleon carried by
the hadronic state $j$.  The Bose statistics of the meson fields
require the splitting functions to be even functions of $y$,
	$f_j(-y) = f_j(y)$,
so that the moments vanish,
	$f_j^{(n)} = 0$,
for all even values of $n = 2, 4, 6 \ldots$ \cite{Chen02}.
From the definition of the PDF moments in Eq.~(\ref{eq:mom_def})
and the crossing symmetry of the quark and antiquark PDFs,
one can further write
\begin{eqnarray}
\langle x^{n-1} \rangle_{q -{\bar q}} 
&=& \Big( 1 - (-1)^n \Big) \langle x^{n-1} \rangle_q\, ,
\label{eq:qminusqbar}
\end{eqnarray}
which implies that for the $q - \bar q$ difference the moments vanish,
$\langle x^{n-1} \rangle_{q - {\bar q}} = 0$, for all even $n$.
Indeed, the matching equation (\ref{eq:match}) can be written
in terms of the moments as
\begin{eqnarray}
\langle x^{n-1} \rangle_{q-\bar q}
&=& \sum_j f_j^{(n)}\, \langle x^{n-1} \rangle_{q/j}\, ,
\label{eq:match_mom}
\end{eqnarray}
with both sides vanishing for $n$ even.
The trivial equality for even $n$ can be removed by limiting the
integration range of the splitting functions $f_j(y)$ to the
physical region between $y=0$ and $y=1$.
To do this, we can define the ``truncated'' moments
${\tilde f}_j^{(n)}$ for physical values of $y$ by
\begin{eqnarray}
{\tilde f}_j^{(n)}
&=& \int_{0}^{1} dy\, y^{n-1} f_j(y),
\label{eq:fbarjn}
\end{eqnarray}
so that
  $f_j^{(n)} = \big( 1-(-1)^n \big) {\tilde f}_j^{(n)}$
by the crossing symmetry property of $f_j(y)$.
Removing the prefactor $\big( 1-(-1)^n \big)$ from both sides
of Eq.~(\ref{eq:match_mom}), one then obtains
\begin{eqnarray}
\langle x^{n-1} \rangle_q 
&=& \sum_j {\tilde f}_j^{(n)}\, \langle x^{n-1} \rangle_{q/j}\, .
\label{eq:match_mom_effective}
\end{eqnarray}
Changing the order of the integrations in ${\tilde f}_j^{(n)}$ and
$\langle x^{n-1} \rangle_{q/j}$, one can write the right-hand side
of Eq.~(\ref{eq:match_mom_effective}) as
\begin{eqnarray}
\sum_j {\tilde f}_j^{(n)}\, \langle x^{n-1} \rangle_{q/j}
&=& \int_{-1}^{1} dx\, x^{n-1} \sum_j \int_0^1 dy\, f_j(y) \int_0^1 dz\, \delta(x-yz)\big( q_j(z)-{\bar q}_j(z) \big),
\end{eqnarray}
so that the left-hand side of (\ref{eq:match_mom_effective})
is equal to
\begin{eqnarray}
\int_{-1}^{1} dx\, x^{n-1} q(x) 
&=& \int_{-1}^{1} dx\, x^{n-1} \sum_j 
    \int_0^1 dy\, f_j(y)
    \int_0^1 dz\, \delta(x-yz)\, q_j^v(z),
\label{eq:convolution_proof}
\end{eqnarray}
where $q_j^v \equiv q_j - \bar{q}_j$ is the valence distribution
for quark flavor $q$ in the hadron $j$.
Since Eq.~(\ref{eq:convolution_proof}) is satisfied for all $n$,
the $x$-integrands of Eqs.~(\ref{eq:match_mom}) and
(\ref{eq:convolution_proof}) must be equivalent, which leads to
the convolution formula for the PDFs,
\begin{eqnarray}  
q(x)
&=& \sum_j \big( f_j \otimes q_j^v \big)(x)\
\equiv\ \sum_j \int_0^1 dy \int_0^1 dz\, \delta(x-yz)\,
               f_j(y)\, q_j^v(z).
\label{eq:conv}
\end{eqnarray}
The convolution expression (\ref{eq:conv}) is the standard one used
in calculations of chiral loop corrections in meson cloud models;
its appearance in the effective chiral theory is made manifest here.

\subsection{Twist-two quark operators}
\label{ssec:twist-2}

From the lowest-order interaction Lagrangians in Eqs.~(\ref{eq:LpBB})
and (\ref{eq:LppBB}), one can derive a set of hadronic operators with
the symmetry properties corresponding to those of the local twist-two
operators in Eq.~(\ref{eq:Oq}).  Specifically, for each quark flavor
$q$, the quark operators can be written in terms of the hadronic
operators according to
\begin{eqnarray}
\label{eq:Oq-hadronic}
\mathcal{O}^{\mu_1 \cdots \mu_n}_q
&=& a^{(n)} i^n \frac{f_\phi^2}{4}
\Big\{
  \mathrm{Tr}
  \left[
    U^{\dag} \lambda^q_+ \partial_{\mu_1} \cdots \partial_{\mu_n} U
  \right]
+ \mathrm{Tr}
  \left[
    U \lambda^q_+ \partial_{\mu_1} \cdots \partial_{\mu_n} U^{\dag}
  \right]
\Big\}							\nonumber\\
&+&
\Big[ \alpha^{(n)} (\overline{{\cal B}} \gamma^{\mu_1} {\cal B} \lambda^q_+)
    + \beta^{(n)}  (\overline{{\cal B}} \gamma^{\mu_1} \lambda^q_+ {\cal B})
    + \sigma^{(n)} (\overline{{\cal B}} \gamma^{\mu_1} {\cal B})\,
      \mathrm{Tr}[\lambda^q_+]
\Big]\, p^{\mu_2} \cdots p^{\mu_n}			\nonumber\\
&+&
\Big[ \bar{\alpha}^{(n)} (\overline{{\cal B}} \gamma^{\mu_1}\gamma_5 {\cal B} \lambda^q_-)
    + \bar{\beta}^{(n)}  (\overline{{\cal B}} \gamma^{\mu_1}\gamma_5 \lambda^q_- {\cal B})
    + \bar{\sigma}^{(n)} (\overline{{\cal B}} \gamma^{\mu_1}\gamma_5 {\cal B})\,
      \mathrm{Tr}[\lambda^q_-]
\Big]\, p^{\mu_2} \cdots p^{\mu_n}			\nonumber\\
&+& \mathrm{permutations} - \mathrm{Tr},
\end{eqnarray}
with a set of {\it a priori} unknown coefficients
  $a^{(n)}$
(for the purely mesonic operators),
  $\{ \alpha^{(n)}, \beta^{(n)}, \sigma^{(n)} \}$
(for the baryonic vector operators), and
  $\{ \bar{\alpha}^{(n)}, \bar{\beta}^{(n)}, \bar{\sigma}^{(n)} \}$
(for the baryonic axial vector operators) for each $n$,
and ``$\mathrm{Tr}$'' represents the trace over the Lorentz indices.
Here, the operator ${\cal B}$ creates spin-1/2 octet baryons,
and the three-index tensor representation of ${\cal B}$ is related
to the octet baryon field matrix $B$ by
\begin{eqnarray}
\label{eq:Octet_representation}
{\cal B}_{ijk}
&=& \frac{1}{\sqrt{6}}
    \left( \epsilon_{ijk'} B^{k'}_k + \epsilon_{ikk'} B^{k'}_j
    \right),
\end{eqnarray}
with the corresponding conjugate representation
\begin{eqnarray}
\label{eq:Octet_conj_representation}
\overline{\cal B}_{kji}
&=& \frac{1}{\sqrt{6}}
    \left( \epsilon_{ijk'} \bar{B}^{k'}_k
	 + \epsilon_{ikk'} \bar{B}^{k'}_j
    \right).
\end{eqnarray}
The flavor operator $\lambda^q_{\pm}$ in Eq.~(\ref{eq:Oq-hadronic})
is defined as
\begin{eqnarray}
\lambda^q_{\pm}
&=& \frac{1}{2}
    \left( u\bar{\lambda}^q u^{\dag} \pm u^{\dag} \bar{\lambda}^q u
    \right),
\end{eqnarray}
with the $3 \times 3$ diagonal matrices $\bar{\lambda}^q$ given by
\begin{eqnarray}
\bar{\lambda}^q
&=& {\rm diag}(\delta_{qu}, \delta_{qd}, \delta_{qs}).
\end{eqnarray}
Expanding up to ${\cal O}(\phi^2)$, this can be written as
\begin{subequations}
\label{eq:lambdaq}
\begin{eqnarray}
\lambda^q_+
&=& \bar{\lambda}^q
 +  \frac{1}{4 f_\phi^2}
    \Big( 2\phi \bar{\lambda}^q \phi - \phi^2 \bar{\lambda}^q
	- \bar{\lambda}^q \phi^2
    \Big)
 +  {\cal O}\left(\phi^4\right),		\\
\lambda^q_-
&=& \frac{i}{\sqrt{2} f_\phi}
    \Big( \phi \bar{\lambda}^q -\bar{\lambda}^q \phi
    \Big) 
 +  {\cal O}\left(\phi^3\right).
\end{eqnarray}
\end{subequations}
%
%
%
The parentheses $(\overline{\cal B} \cdots {\cal B})$ in
(\ref{eq:Oq-hadronic}), involving the three-index tensor
representation of the ${\cal B}$ operator, are related to the
ordinary traces of the baryon field matrix $B$ using the
identities~\cite{Labrenz:1996}
\begin{subequations}
\begin{eqnarray}
(\overline{\cal B} {\cal B})
&=& \mathrm{Tr}\big[\bar{B}B\big],			\\
(\overline{\cal B} {\cal B} A)
&=& \frac{2}{3}\mathrm{Tr}\big[\bar{B} A B\big]
 +  \frac{1}{6}\mathrm{Tr}\big[\bar{B}B\big] \mathrm{Tr}\big[A\big]
 -  \frac{1}{6}\mathrm{Tr}\big[\bar{B} B A\big],	\\
(\overline{\cal B} A {\cal B})
&=& -\frac{1}{3}\mathrm{Tr}\big[\bar{B} A B\big]
 +  \frac{2}{3}\mathrm{Tr}\big[\bar{B}B\big] \mathrm{Tr}\big[A\big]
 -  \frac{2}{3}\mathrm{Tr}\big[\bar{B} B A\big].
\end{eqnarray}
\end{subequations}
Using these relations, the hadronic operators for the $u$ and $d$
quark flavors relevant to the nucleon initial and final states
can be expanded as
\begin{eqnarray}\label{eq:Ou}
\mathcal{O}^{\mu_1 \cdots \mu_n}_u 
&=& \frac{a^{(n)}}{2}
    \left( \mathcal{O}^{\mu_1 \cdots \mu_n}_{\pi^+}
	 + \mathcal{O}^{\mu_1 \cdots \mu_n}_{K^+}
    \right)						\nonumber\\
&+& 
\Big[ 
\Big( \frac{5}{6}\alpha^{(n)}
    + \frac{1}{3}\beta^{(n)}
    + \sigma^{(n)}
\Big) \mathcal{O}^{\mu_1 \cdots \mu_n}_{p}
+
\Big( \frac{1}{6}\alpha^{(n)}
    + \frac{2}{3}\beta^{(n)}
    + \sigma^{(n)}
\Big) \mathcal{O}^{\mu_1 \cdots \mu_n}_{n}  		\nonumber\\
&&
+
\Big( \frac{1}{6}\alpha^{(n)}
    + \frac{2}{3}\beta^{(n)}
    + \sigma^{(n)}
\Big) \mathcal{O}^{\mu_1 \cdots \mu_n}_{\Xi^0} 
+
\Big( \frac{1}{4}\alpha^{(n)}
    + \frac{1}{2}\beta^{(n)}
    + \sigma^{(n)}
\Big) \mathcal{O}^{\mu_1 \cdots \mu_n}_{\Lambda}	\nonumber\\
&&
+
\Big( \frac{5}{12}\alpha^{(n)}
    + \frac{1}{6}\beta^{(n)}
    + \sigma^{(n)}
\Big) \mathcal{O}^{\mu_1 \cdots \mu_n}_{\Sigma^0} 
+
\Big( \frac{5}{6}\alpha^{(n)}
    + \frac{1}{3}\beta^{(n)}
    + \sigma^{(n)}
\Big) \mathcal{O}^{\mu_1 \cdots \mu_n}_{\Sigma^+} 	\nonumber\\
&&
+ \sigma^{(n)}
  \Big( \mathcal{O}^{\mu_1 \cdots \mu_n}_{\Sigma^-}
      + \mathcal{O}^{\mu_1 \cdots \mu_n}_{\Xi^-}
  \Big)
+
\frac{1}{4\sqrt{3}}
\Big( \alpha^{(n)}
     - 2 \beta^{(n)}
\Big)
\Big( \mathcal{O}^{\mu_1 \cdots \mu_n}_{\Lambda \Sigma^0} 
    + \mathcal{O}^{\mu_1 \cdots \mu_n}_{\Sigma^0 \Lambda}
\Big)
\Big]							\nonumber\\
&+& 
\Big[
  \frac{1}{12} (-4 \alpha^{(n)} + 2 \beta^{(n)})
  \mathcal{O}^{\mu_1 \cdots \mu_n}_{\bar{p}p\pi^+ \pi^-}
+ \frac{1}{12} ( -5 \alpha^{(n)} - 2 \beta^{(n)})
  \mathcal{O}^{\mu_1 \cdots \mu_n}_{\bar{p}pK^+ K^-}	\nonumber\\
&&
+ \frac{1}{12} ( 4 \alpha^{(n)} - 2 \beta^{(n)})
  \mathcal{O}^{\mu_1 \cdots \mu_n}_{\bar{n}n\pi^+ \pi^-}
+ \frac{1}{12} ( - \alpha^{(n)} - 4 \beta^{(n)})
  \mathcal{O}^{\mu_1 \cdots \mu_n}_{\bar{n}nK^+ K^-}
\Big]                   				\nonumber\\ 
&+&
\frac{1}{3\sqrt{2}}
\Big( 2 \bar{\alpha}^{(n)} - \bar{\beta}^{(n)}
\Big) \mathcal{O}^{\mu_1 \cdots \mu_n}_{n p \pi^-}
-
\frac{\sqrt{3}}{4} \bar{\alpha}^{(n)}
\mathcal{O}^{\mu_1 \cdots \mu_n}_{\Lambda p K^-}	\nonumber\\
&-&
\frac{1}{12}
\Big( \bar{\alpha}^{(n)} + 4 \bar{\beta}^{(n)}
\Big)
\Big( \mathcal{O}^{\mu_1 \cdots \mu_n}_{\Sigma^0 p K^-}
  + \sqrt{2} \mathcal{O}^{\mu_1 \cdots \mu_n}_{\Sigma^- n K^-}
\Big),
\end{eqnarray}
and
\begin{eqnarray}\label{eq:Od}
\mathcal{O}^{\mu_1 \cdots \mu_n}_d 
&=& \frac{a^{(n)}}{2}
\Big( \mathcal{O}^{\mu_1 \cdots \mu_n}_{\pi^-}
    + \mathcal{O}^{\mu_1 \cdots \mu_n}_{K^0}
\Big)							\nonumber\\
&+& 
\Big[
\Big( \frac{1}{6}\alpha^{(n)}
    + \frac{2}{3}\beta^{(n)}
    + \sigma^{(n)}
\Big) \mathcal{O}^{\mu_1 \cdots \mu_n}_{p}
+
\Big( \frac{5}{6}\alpha^{(n)}
    + \frac{1}{3}\beta^{(n)}
    + \sigma^{(n)}
\Big) \mathcal{O}^{\mu_1 \cdots \mu_n}_{n}		\nonumber\\
&& 
+
\Big( \frac{1}{6}\alpha^{(n)}
      + \frac{2}{3}\beta^{(n)}
      + \sigma^{(n)}
\Big) \mathcal{O}^{\mu_1 \cdots \mu_n}_{\Xi^-} 
+
\Big( \frac{1}{4}\alpha^{(n)}
    + \frac{1}{2}\beta^{(n)}
    + \sigma^{(n)}
\Big) \mathcal{O}^{\mu_1 \cdots \mu_n}_{\Lambda}	\nonumber\\
&&
+
\Big( \frac{5}{12}\alpha^{(n)}
      + \frac{1}{6}\beta^{(n)}
      + \sigma^{(n)}
\Big) \mathcal{O}^{\mu_1 \cdots \mu_n}_{\Sigma^0} 
+
\Big( \frac{5}{6}\alpha^{(n)}
    + \frac{1}{3}\beta^{(n)}
    + \sigma^{(n)}
\Big) \mathcal{O}^{\mu_1 \cdots \mu_n}_{\Sigma^-} 	\nonumber\\
&&
+ \sigma^{(n)}
  \Big( \mathcal{O}^{\mu_1 \cdots \mu_n}_{\Sigma^+}
      + \mathcal{O}^{\mu_1 \cdots \mu_n}_{\Xi^0}
  \Big)
-
\frac{1}{4\sqrt{3}}
\Big( \alpha^{(n)}
     - 2 \beta^{(n)}
\Big)
\Big( \mathcal{O}^{\mu_1 \cdots \mu_n}_{\Lambda \Sigma^0} 
    + \mathcal{O}^{\mu_1 \cdots \mu_n}_{\Sigma^0 \Lambda}
\Big)
\Big]							\nonumber\\
&+& 
\Big[
  \frac{1}{12} (4 \alpha^{(n)} - 2 \beta^{(n)})
  \mathcal{O}^{\mu_1 \cdots \mu_n}_{\bar{p}p\pi^+ \pi^-}
+ \frac{1}{12} ( - \alpha^{(n)} - 4 \beta^{(n)})
  \mathcal{O}^{\mu_1 \cdots \mu_n}_{\bar{p}pK^0 \bar{K}^0}  \nonumber\\
&&
+ \frac{1}{12} ( - 4 \alpha^{(n)} + 2 \beta^{(n)})
  \mathcal{O}^{\mu_1 \cdots \mu_n}_{\bar{n}n\pi^+ \pi^-}
+ \frac{1}{12} ( - 5 \alpha^{(n)} - 2 \beta^{(n)})
  \mathcal{O}^{\mu_1 \cdots \mu_n}_{\bar{n}nK^0 \bar{K}^0}
\Big]                   				\nonumber\\ 
&-& \frac{1}{3\sqrt{2}}
\Big( 2 \bar{\alpha}^{(n)} - \bar{\beta}^{(n)}
\Big) \mathcal{O}^{\mu_1 \cdots \mu_n}_{n p \pi^-}
-
\frac{\sqrt{3}}{4} \bar{\alpha}^{(n)}
\mathcal{O}^{\mu_1 \cdots \mu_n}_{\Lambda n \bar{K}^0}	\nonumber\\
&-& \frac{1}{12}
\Big( \bar{\alpha}^{(n)} + 4 \bar{\beta}^{(n)}
\Big)
\Big(
 - \mathcal{O}^{\mu_1 \cdots \mu_n}_{\Sigma^0 n \bar{K}^0}
 + \sqrt{2} \mathcal{O}^{\mu_1 \cdots \mu_n}_{\Sigma^+ p \bar{K}^0}
\Big),
\end{eqnarray}
respectively.  For the twist-two strange quark operator, which is
directly relevant to the current analysis, one has
\begin{eqnarray}
\label{eq:Os}
\mathcal{O}^{\mu_1 \cdots \mu_n}_s
&=& -\frac{a^{(n)}}{2}
    \left( \mathcal{O}^{\mu_1 \cdots \mu_n}_{K^+}
	 + \mathcal{O}^{\mu_1 \cdots \mu_n}_{K^0}
    \right)						\nonumber\\
&+&
\Big[
\Big( \frac{1}{2} \alpha^{(n)} + \sigma^{(n)}
\Big) \mathcal{O}^{\mu_1 \cdots \mu_n}_{\Lambda}
+
\Big( \frac{1}{6}\alpha^{(n)}
    + \frac{2}{3}\beta^{(n)}
    + \sigma^{(n)}
\Big) 
\Big( \mathcal{O}^{\mu_1 \cdots \mu_n}_{\Sigma^+}
    + \mathcal{O}^{\mu_1 \cdots \mu_n}_{\Sigma^0}
    + \mathcal{O}^{\mu_1 \cdots \mu_n}_{\Sigma^-}
\Big)							\nonumber\\
&&
+
\Big( \frac{5}{6}\alpha^{(n)}
    + \frac{1}{3}\beta^{(n)}
    + \sigma^{(n)}
\Big)
\Big( \mathcal{O}^{\mu_1 \cdots \mu_n}_{\Xi^-}
    + \mathcal{O}^{\mu_1 \cdots \mu_n}_{\Xi^0}
\Big)
+
\sigma^{(n)}
\Big( \mathcal{O}^{\mu_1 \cdots \mu_n}_{p}
    + \mathcal{O}^{\mu_1 \cdots \mu_n}_{n}
\Big)
\Big]							\nonumber\\
&+& \frac{1}{12} (5 \alpha^{(n)} + 2 \beta^{(n)})
\Big(
  \mathcal{O}^{\mu_1 \cdots \mu_n}_{\bar{p}pK^+ K^-}
+ \mathcal{O}^{\mu_1 \cdots \mu_n}_{\bar{n}nK^0 \bar{K}^0}
\Big)
+
\frac{1}{12} (\alpha^{(n)} + 4 \beta^{(n)})
\Big( \mathcal{O}^{\mu_1 \cdots \mu_n}_{\bar{p}pK^0\bar{K}^0}
    + \mathcal{O}^{\mu_1 \cdots \mu_n}_{\bar{n}nK^+K^-}
\Big)							\nonumber\\
&+&
\frac{1}{6} (2\alpha^{(n)} - \beta^{(n)})
\Big( \mathcal{O}^{\mu_1 \cdots \mu_n}_{\bar{p}nK^+\bar{K}^0}
    + \mathcal{O}^{\mu_1 \cdots \mu_n}_{\bar{n}pK^0K^-}
\Big)							\nonumber\\
&+&
\frac{\sqrt{3}}{4} \bar{\alpha}^{(n)}
\Big( \mathcal{O}^{\mu_1 \cdots \mu_n}_{\Lambda p K^-}
    + \mathcal{O}^{\mu_1 \cdots \mu_n}_{\Lambda n \bar{K}^0}
\Big)
+
\frac{1}{12}
\Big( \bar{\alpha}^{(n)} + 4 \bar{\beta}^{(n)}
\Big)
\Big( \mathcal{O}^{\mu_1 \cdots \mu_n}_{\Sigma^0 p K^-}
   + \sqrt{2} \mathcal{O}^{\mu_1 \cdots \mu_n}_{\Sigma^+ p \bar{K}^0}
\Big)							\nonumber\\
&+& \frac{1}{12}
\Big( \bar{\alpha}^{(n)} + 4 \bar{\beta}^{(n)}
\Big)
\Big(
 - \mathcal{O}^{\mu_1 \cdots \mu_n}_{\Sigma^0 n \bar{K}^0}
 + \sqrt{2} \mathcal{O}^{\mu_1 \cdots \mu_n}_{\Sigma^- n K^-}
\Big).
\end{eqnarray}
The various hadronic operators in Eqs.~(\ref{eq:Ou}) -- (\ref{eq:Os})
are defined as
\begin{subequations}
\label{eq:had_ops}
\begin{eqnarray}
\mathcal{O}^{\mu_1 \cdots \mu_n}_\phi
&=& i^n
    \big( \bar{\phi}\, \partial^{\mu_1} \cdots \partial^{\mu_n} \phi
        - \phi\, \partial^{\mu_1} \cdots \partial^{\mu_n} \bar{\phi}
    \big),						\\
\mathcal{O}^{\mu_1 \cdots \mu_n}_{B' B}
&=& \big( \bar{B}' \gamma^{\mu_1} B \big)\,
    p^{\mu_2} \cdots p^{\mu_n},
\label{eq:OB'B}						\\
\mathcal{O}^{\mu_1 \cdots \mu_n}_{B B \phi \phi}
&=& \frac{1}{f^2_\phi}
    \big( \bar{B} \gamma^{\mu_1} B \bar{\phi}\, \phi \big)\,
    p^{\mu_2} \cdots p^{\mu_n},				\\
\mathcal{O}^{\mu_1 \cdots \mu_n}_{B'B\phi}
&=& \frac{i}{f_\phi}
    \big( \bar{B'} \gamma^{\mu_1}\gamma_5 B \phi
        - \bar{B}  \gamma^{\mu_1}\gamma_5 B' \bar{\phi}
    \big)\,
    p^{\mu_2} \cdots p^{\mu_n},
\label{eq:OB'BP}
\end{eqnarray}
\end{subequations}
where for the $B' B$ and $B'B\phi$ operators in Eqs.~(\ref{eq:OB'B})
and (\ref{eq:OB'BP}) the fields $B$ and $B'$ can in principle be
different.

\begin{figure}[t]
\includegraphics[width=4.0in]{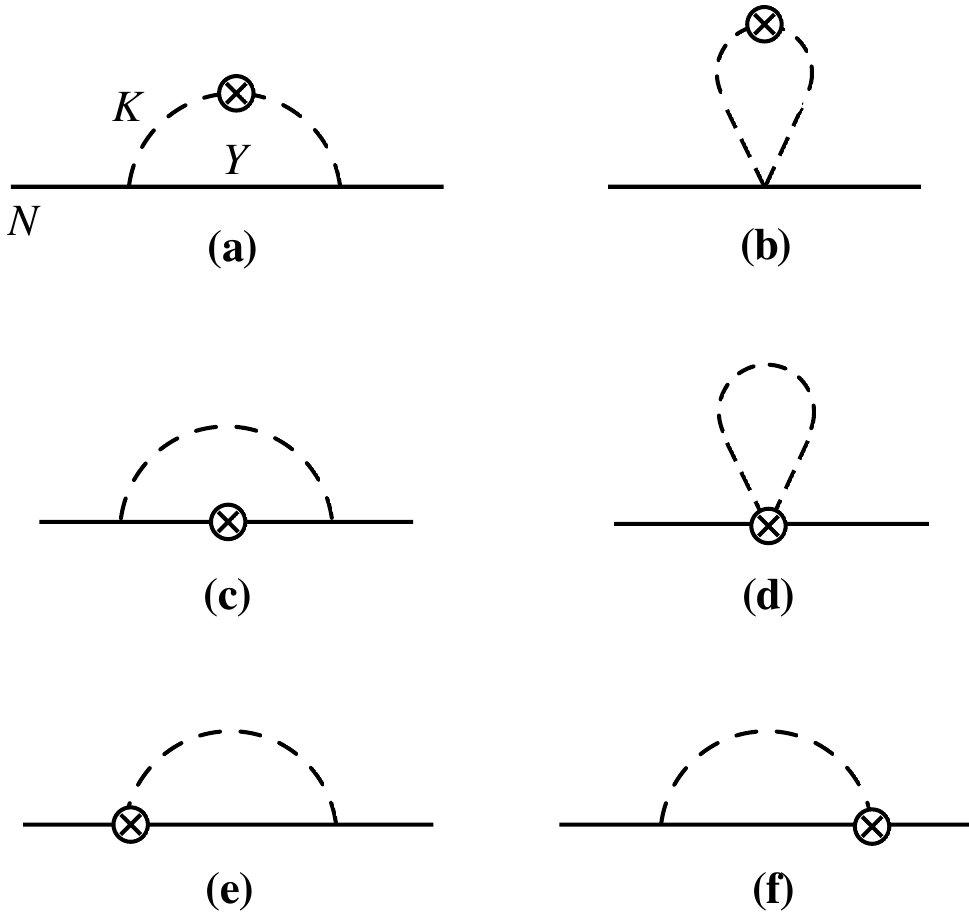}
\caption{Contributions to the $\bar s$ PDF in the nucleon from
	(a) the kaon rainbow and
	(b) kaon bubble diagrams,
	and contributions to the $s$ PDF from
	(c) the hyperon rainbow,
	(d) kaon tadpole, and
	(e), (f) Kroll-Ruderman diagrams.
	The kaons, $K$, and hyperon, $Y$, are represented by
	the internal dashed and solid curves, respectively,
	and the crosses represent insertions of the operators
	in Eq.~(\ref{eq:Os}).}
\label{fig:loops}
\end{figure}

From the operator structures in Eq.~(\ref{eq:Os}) we can identify
several distinct contributions to the nucleon matrix elements of
the strange quark twist-two operators.
These are illustrated in Fig.~\ref{fig:loops}, and include the
kaon and hyperon rainbow diagrams, the kaon bubble and tadpole
contributions, and the Kroll-Ruderman terms that are necessary
for the preservation of gauge invariance.
Each of these can be expressed in terms of a particular
nucleon $\to$ strange hadron splitting function $f_j(y)$
and the corresponding PDF in the strange hadron.
The moments of the latter can be related to various
combinations of coefficients of the hadronic operators
in Eq.~(\ref{eq:Os}), as we discuss next.

\subsection{Matching coefficients and PDF moments}
\label{ssec:match_coeff}

Generally, the coefficients of the operators in
Eq.~(\ref{eq:Oq-hadronic}) are not constrained by symmetries
and must be determined from elsewhere.  Within the convolution
formalism, Eq.~(\ref{eq:conv}), the coefficient $a^{(n)}$ is
related, for example, to the $u$-quark or $\bar s$-antiquark
distribution in the $K^+$ meson,
\begin{equation}
\frac{a^{(n)}}{2} = \int_{-1}^1 dx\, x^{n-1}\, \bar s_{K^+}(x),
\label{eq:an}
\end{equation}
from which we have $a^{(1)} = 2$.
Within the chiral SU(3) framework, the kaon and pion PDFs are
related by
$\bar s_{K^+} = u_{K^+} = \bar s_{K^0}
= u_{\pi^+} = \bar{d}_{\pi^+} = d_{\pi^-} = \bar{u}_{\pi^-}$
for all $x$ values.

The coefficients $\alpha^{(n)}$, $\beta^{(n)}$ and $\sigma^{(n)}$,
on the other hand, are related to the moments of the $u$, $d$ and $s$
PDFs in the bare proton,
\begin{subequations}
\begin{eqnarray}
\frac{5}{6} \alpha^{(n)} + \frac{1}{3} \beta^{(n)} + \sigma^{(n)}
&=& \int_{-1}^{1} dx\, x^{n-1}\, u(x),		\\
\frac{1}{6} \alpha^{(n)} + \frac{2}{3} \beta^{(n)} + \sigma^{(n)}
&=& \int_{-1}^{1} dx\, x^{n-1}\, d(x),		\\
\sigma^{(n)}
&=& \int_{-1}^{1} dx\, x^{n-1}\, s(x).
\label{eq:sigma}
\end{eqnarray}
\label{eq:alpha-beta-sigma-original}
\end{subequations}
Solving Eqs.~(\ref{eq:alpha-beta-sigma-original}), these coefficients
can be obtained in terms of the proton PDFs,
\begin{subequations}  
\label{eq:alpha-beta-sigma}
\begin{eqnarray}
\alpha^{(n)}
&=& \int_{-1}^1 dx\, x^{n-1}
    \Big(  \frac{4}{3} u(x) - \frac{2}{3} d(x) - \frac{2}{3} s(x) \Big), \\
\beta^{(n)}
&=& \int_{-1}^1 dx\, x^{n-1}
    \Big( -\frac{1}{3} u(x) + \frac{5}{3} d(x) - \frac{4}{3} s(x) \Big),
%
\end{eqnarray}
\end{subequations}%
with $\sigma^{(n)}$ given by Eq.~(\ref{eq:sigma}).
Note that in the SU(3) symmetric limit, the strange quark PDF $s(x)$
in the nucleon is identically zero; for the time being, we keep it
explicitly in Eqs.~(\ref{eq:alpha-beta-sigma}) for generality.
For $n=1$, the coefficients are then fixed by the conservation of
the total charge and strangeness in the nucleon,
\begin{equation}
\alpha^{(1)} = 2,\ \ \ \
\beta^{(1)}  = 1,\ \ \ \
\sigma^{(1)} = 0.
\label{eq:coeff_norm}
\end{equation}

To determine the coefficients $\bar{\alpha}^{(n)}$, $\bar{\beta}^{(n)}$
and $\bar{\sigma}^{(n)}$ of the axial vector operators, in contrast,
one needs to consider spin-dependent twist-two operators,
\begin{equation}\label{eq:O-delta-q}
\mathcal{O}^{\mu_1 \cdots \mu_n}_{\Delta q}
= i^{n-1}
  \bar{q} \gamma_5 \gamma^{ \{\mu_1 }
  \overleftrightarrow{D}^{\mu_2} \cdots
  \overleftrightarrow{D}^{ \mu_n\} } q.
\end{equation}
In the effective field theory the spin-dependent twist-two operators
can be matched to the hadronic operators according to \cite{Shanahan13}
\begin{eqnarray}
\label{eq:O-delta-q-hadronic}
\mathcal{O}^{\mu_1 \cdots \mu_n}_{\Delta q}
&=&
\Big[
  \bar{\alpha}^{(n)}(\overline{\cal B} \gamma^{\mu_1}\gamma_5 {\cal B}\lambda^q_+)
+ \bar{\beta}^{(n)}(\overline{\cal B} \gamma^{\mu_1}\gamma_5 \lambda^q_+ {\cal B})
+ \bar{\sigma}^{(n)}(\overline{\cal B} \gamma^{\mu_1}\gamma_5 {\cal B})\,
  \mathrm{Tr}[\lambda^q_+]
\Big] p^{\mu_2} \cdots p^{\mu_n}			\nonumber\\
&+&
\Big[
  \alpha^{(n)}(\overline{\cal B} \gamma^{\mu_1} {\cal B} \lambda^q_-)
+ \beta^{(n)}(\overline{\cal B} \gamma^{\mu_1} \lambda^q_- {\cal B})
+ \sigma^{(n)}(\overline{\cal B} \gamma^{\mu_1} {\cal B})\,
  \mathrm{Tr}[\lambda^q_-]
\Big] p^{\mu_2} \cdots p^{\mu_n}			\nonumber\\
&+& \mathrm{permutations}\ -\ \mathrm{Tr}.
\end{eqnarray}
According to the properties of $\mathcal{O}^{\mu_1 \cdots \mu_n}_{q}$
and $\mathcal{O}^{\mu_1 \cdots \mu_n}_{\Delta q}$ under parity
transformations \cite{Moiseeva:2013}, the coefficients
  $\{ \alpha^{(n)}, \beta^{(n)}, \sigma^{(n)} \}$ and 
  $\{ \bar{\alpha}^{(n)}, \bar{\beta}^{(n)}, \bar{\sigma}^{(n)} \}$
are the same as for the spin-averaged operators in
Eq.~(\ref{eq:Oq-hadronic}).
Expanding Eq.~(\ref{eq:O-delta-q-hadronic}) to lowest order, the
coefficients can then be related to the moments of the spin-dependent
PDFs in the bare proton by
\begin{subequations}
\begin{eqnarray}
\frac{5}{6} \bar{\alpha}^{(n)}
+ \frac{1}{3} \bar{\beta}^{(n)}
+ \bar{\sigma}^{(n)}
&=& \int_{-1}^{1} dx\, x^{n-1} \Delta u(x),	\\
\frac{1}{6} \bar{\alpha}^{(n)}
+ \frac{2}{3} \bar{\beta}^{(n)}
+ \bar{\sigma}^{(n)}
&=& \int_{-1}^{1} dx\, x^{n-1} \Delta d(x),	\\
\bar{\sigma}^{(n)}
&=& \int_{-1}^{1} dx\, x^{n-1} \Delta s(x),
\end{eqnarray}
\end{subequations}
from which the individual coefficients can be determined according to
\begin{subequations}
\begin{eqnarray}
\bar{\alpha}^{(n)}
&=& \int_{-1}^1 dx\, x^{n-1}
    \left( \frac{4}{3} \Delta u(x)
	 - \frac{2}{3} \Delta d(x)
	 - \frac{2}{3} \Delta s(x)
    \right),					\\
\bar{\beta}^{(n)}
&=& \int_{-1}^1 dx\, x^{n-1}
    \left(-\frac{1}{3} \Delta u(x)
	 + \frac{5}{3} \Delta d(x)
	 - \frac{4}{3} \Delta s(x)
    \right).
%
\end{eqnarray}
\label{eq:alpha-beta-sigma-bar}
\end{subequations}  
As for the spin-averaged PDF in Eqs.~(\ref{eq:alpha-beta-sigma}),
here we again keep the bare polarized strange quark PDF $\Delta s(x)$
in the nucleon for generality, even though in the SU(3) limit it is
zero.  For $n=1$, the coefficients $\bar{\alpha}^{(1)}$ and
$\bar{\beta}^{(1)}$ are fixed from the SU(3) decay constants by
\begin{equation}
\bar{\alpha}^{(1)} = \frac{2}{3}(D + 3F),\ \ \ \
\bar{\beta}^{(1)}  = - \frac{1}{3} (5D - 3F).
\label{eq:barcoeff_norm}
\end{equation}

Along with the nucleon and meson PDFs that appear the calculation
of the PDFs in Eq.~(\ref{eq:conv}), contributions from PDFs in
strange baryons also enter the convolution integrals.
Within the chiral SU(3) framework, moments of the strange quark PDFs
in the hyperons, $s_Y$, are given in terms of the coefficients by
\begin{subequations}
\begin{eqnarray}
\frac{1}{2}\alpha^{(n)} + \sigma^{(n)}
&=& \int_{-1}^1 dx\, x^{n-1} s_{\Lambda}(x),	\\
\frac{1}{6}\alpha^{(n)} + \frac{2}{3}\beta^{(n)} + \sigma^{(n)}
&=& \int_{-1}^1 dx\, x^{n-1} s_{\Sigma^+}(x)\
 =\ \int_{-1}^1 dx\, x^{n-1} s_{\Sigma^0}(x).
\end{eqnarray}
\end{subequations}
Combining with Eqs.~(\ref{eq:alpha-beta-sigma}), the strange PDFs
in the $\Lambda$ and $\Sigma$ hyperons are then related to the
$u$ and $d$ PDFs in the proton according to
\begin{subequations}
\label{eq:sY}
\begin{eqnarray}
s_{\Lambda}(x)
&=& \frac{1}{3} \Big[ 2u(x) - d(x) + 2s(x) \Big],
\label{eq:slambda}				\\
s_{\Sigma^+}(x)
&=& s_{\Sigma^0}(x) = d(x).
\label{eq:ssigma}
\end{eqnarray}
\end{subequations}
In practice SU(3) symmetry violating effects \cite{Bass10} may give
corrections to these relations at the 10\%--20\% level \cite{XWang16},
although a dedicated study of the phenomenological impact on PDFs
will be necessary for a more quantitative estimate.

For the strange PDFs associated with the Kroll-Ruderman vertices
in Figs.~\ref{fig:loops}(e) and (f), $s^{(\rm KR)}_Y(x)$, one makes
use of the moment relations
\begin{subequations}
\begin{eqnarray}
\frac{\bar{\alpha}^{(n)}}{\bar{\alpha}^{(1)}}
&=& \int_{-1}^1 dx\, x^{n-1} s^{(\mathrm{KR})}_{\Lambda}(x),	\\
\frac{\bar{\alpha}^{(n)}
 +  4\bar{\beta}^{(n)}}{ \bar{\alpha}^{(1)}
 +  4 \bar{\beta}^{(1)}}
&=& \int_{-1}^1 dx\, x^{n-1} s^{(\mathrm{KR})}_{\Sigma^+}(x)\
 =\ \int_{-1}^1 dx\, x^{n-1} s^{(\mathrm{KR})}_{\Sigma^0}(x).
\end{eqnarray}
\end{subequations}
Combining with Eqs.~(\ref{eq:alpha-beta-sigma-bar}), the Kroll-Ruderman
strange-quark distributions can then be written in terms of
spin-dependent PDFs in the nucleon,
\begin{subequations}
\label{eq:sKR}
\begin{eqnarray}
s^{(\mathrm{KR})}_{\Lambda}(x)
&=& \frac{1}{D+3F}\, \Big[ 2 \Delta u(x) - \Delta d(x) \Big],	\\
s^{(\mathrm{KR})}_{\Sigma^+}(x)
&=& s^{(\mathrm{KR})}_{\Sigma^0}(x) = \frac{1}{F-D}\, \Delta d(x).
\end{eqnarray}
\end{subequations}

Finally, for the strange quark distributions relevant for the
Weinberg-Tomozawa tadpole contribution in Fig.~\ref{fig:loops}(d),
$s_K^{\rm (tad)}(y)$, one finds the moment relations
\begin{subequations}
\begin{eqnarray}
\frac{1}{12}\left( 5\alpha^{(n)} + 2\beta^{(n)} \right)
&=& \int_{-1}^1 dx\, x^{n-1}\, s_{K^+}^{\rm (tad)}(x),		\\
\frac{1}{6}\left( \alpha^{(n)} + 4\beta^{(n)} \right)
&=& \int_{-1}^1 dx\, x^{n-1}\, s_{K^0}^{\rm (tad)}(x).
\end{eqnarray}
\end{subequations}
Combining with Eqs.~(\ref{eq:alpha-beta-sigma}), the PDFs associated
with the charged and neutral kaon loops are given by
\begin{subequations}
\label{eq:s-tad}
\begin{eqnarray}
s_{K^+}^{\rm (tad)}(x) &=& \frac{1}{2} u(x),	\\
s_{K^0}^{\rm (tad)}(x) &=& d(x).
\end{eqnarray}
\end{subequations}

These relations provide the complete information on the PDFs in the
strange hadrons necessary for the computation of the loop diagrams of
Fig.~\ref{fig:loops}.  The remaining ingredients needed to evaluate
the convolutions in Eq.~(\ref{eq:conv}) are the hadronic splitting
functions $f_j(y)$.  In the next section we derive these from the
matrix elements of the operators listed in Sec.~\ref{ssec:twist-2}.

\section{Hadronic splitting functions}
\label{sec:fy}

The hadronic splitting functions $f_j(y)$ defined in
Eqs.~(\ref{eq:fjn_def}) and (\ref{eq:fjn}) can be thought of as the
effective theory analogs of the quark and gluon splitting functions
of perturbative QCD that enter in the PDF evolution equations
\cite{Altarelli77}.  In this case the nucleon $\to$ kaon $+$ hyperon
splitting functions are evaluated for each of the hadronic level
diagrams in Fig.~\ref{fig:loops}, with the interaction vertices
given by the operators in Eqs.~(\ref{eq:Os}) and (\ref{eq:had_ops}).
In this section we give the complete set of strange hadron splitting
functions in the effective theory.
Regularization of the functions will be discussed in Sec.~\ref{sec:reg}.
In general we follow the notations introduced for the pion loop
corrections in Refs.~\cite{Z1, Burkardt13, Salamu15, XWang16},
with obvious extensions.

\subsection{Kaon rainbow distribution}
\label{ssec:fyKrbw}

We begin with the light-cone distributions associated with the
operator insertions on the kaon loop.  These give rise to two types
of diagrams, illustrated in Fig.~\ref{fig:loops}, involving the kaon
rainbow and contact interactions.  For the kaon rainbow diagram in
Fig.~\ref{fig:loops}(a), the splitting function is given by
\begin{eqnarray}
\label{eq:kaon-LC-distribution}
f_{KY}^{\rm (rbw)}(y)
&=& M \frac{C_{KY}^2}{f_\phi^2}
    \int\!\frac{d^4 k}{(2\pi)^4}\,
    \bar{u}(p) (\slashed{k} \gamma_5)
    \frac{i(\slashed{p} - \slashed{k} + M_Y)}{D_Y}
    (\gamma_5 \slashed{k}) u(p)
    \frac{i}{D_K} \frac{i}{D_K}
    2 k^+ \delta(k^+ - y p^+),			\nonumber\\
&&
\end{eqnarray}
where $p$ and $k$ are the physical nucleon and virtual kaon
four-momenta, and $D_K$ and $D_Y$ are the kaon and hyperon
virtualities, given by
\begin{subequations}
\begin{eqnarray}
D_K &=& k^2 - m_K^2 + i\epsilon,		\\
D_Y &=& (p-k)^2 - M_Y^2 + i\epsilon,
\end{eqnarray}
\end{subequations}
respectively, with $m_K$ and $M_Y$ the corresponding kaon and
hyperon masses.
The spinors $u(p)$ are normalized such that $\bar{u}(p) u(p) = 1$.
The coefficients $C_{KY}^2$ can be obtained from the effective
Lagrangian in Eq.~(\ref{eq:LpBB}),
\begin{equation}
C_{K^+ \Lambda}^2
= \left( \frac{D+3F}{2\sqrt{3}} \right)^2,\ \ \ \ \
C_{K^0 \Sigma^+}^2
= 2 C_{K^+\Sigma^0}^2
= \left( \frac{D-F}{\sqrt{2}} \right)^2.
\end{equation}
Using the Dirac equation, the integrand in
Eq.~(\ref{eq:kaon-LC-distribution}) can be decomposed into
several terms,
\begin{eqnarray}
\label{eq:rainbow}
f_{KY}^{\rm (rbw)}(y)
&=& - i \frac{C_{KY}^2}{f_\phi^2}
    \int\!\frac{d^4 k}{(2\pi)^4}
    \left[ \frac{\overline{M}^2 (p \cdot k + M \Delta)}{D_K^2 D_Y}
	 + \frac{M \overline{M}}{D^2_K}
	 + \frac{p \cdot k}{D^2_K}
    \right]
    2 y\, \delta\left( y - \frac{k^+}{p^+} \right),
\end{eqnarray}
where the sum and difference of the hyperon and nucleon masses
are defined as
\begin{subequations}
\begin{eqnarray}
\overline{M} &=& M_Y + M,	\\
\Delta &=& M_Y - M,
\end{eqnarray}
\end{subequations}
respectively.
(Note that $\overline{M}$ and $\Delta$ should both have an index
``$Y$'' to differentiate between the $\Lambda$ and $\Sigma$ masses;
for notational convenience, however, we suppress them in the following.)
Using the residue theorem to perform the $k^-$ integration and closing
the contour in the upper half plane to take the hyperon pole,
\begin{equation}
D_Y = (p^+ - k^+)
      \left( p^- - k^-
           - \frac{k^2_{\perp} + M_Y^2 - i \epsilon}{p^+ - k^+}
      \right)\ \to\ 0,
\end{equation}
one can show that the first term ($\sim 1/D_K^2 D_Y$) in the
brackets of Eq.~(\ref{eq:rainbow}) corresponds to the on-shell
hyperon contribution.  This term contributes at $y>0$, and is
the contribution usually associated with the ``Sullivan process''
\cite{Sullivan72, Signal87}.
The second term ($\sim 1/D^2_K$) in Eq.~(\ref{eq:rainbow})
vanishes after integration by symmetry arguments \cite{Burkardt13}.
Using the identity \cite{Z1}
\begin{equation}\label{eq:identity}
\int d^4k \frac{2y\, p\cdot k}{D_K^2}
= \int d^4k \frac{1}{D_K},
\end{equation}
the third term in Eq.~(\ref{eq:rainbow}) can be shown to give
a singular contribution at $y=0$ \cite{Burkardt13}.
The splitting function for the kaon rainbow diagram can then be
written as a sum of the on-shell and contact ($\delta$-function)
contributions,
\begin{equation}
f_{KY}^{(\rm rbw)}(y)
= \frac{C_{KY}^2 \overline{M}^2}{(4\pi f_\phi)^2}
  \left[ f_Y^{\rm (on)}(y) + f_K^{(\delta)}(y)
  \right].
\label{eq:fKYrbw}
\end{equation}     
The on-shell function is given by
\begin{equation}
f_Y^{\rm (on)}(y)
= y \int\!dk_\perp^2\,
  \frac{k_\perp^2 + (M y + \Delta)^2}{(1-y)^2 D_{KY}^2}
  F^{\rm (on)},
\label{eq:fYon}
\end{equation}
where
\begin{eqnarray}
D_{KY}
&=& -\left[ \frac{k_\perp^2 + y M_Y^2 + (1-y) m_K^2 - y(1-y) M^2}{1-y}
     \right]
\label{eq:DKY}
\end{eqnarray}
is the kaon virtuality for an on-shell hyperon intermediate state.
Since the splitting functions for point-particles are ultraviolet
divergent, a regularization prescription needs to be used to obtain
finite results.  Anticipating the discussion of the ultraviolet
regularization in Sec.~\ref{sec:reg} below, we introduce in
Eq.~(\ref{eq:fYon}) a function $F^{\rm (on)}$ that regularizes
the ultraviolet divergence of the $k_\perp^2$ integration.
The expression in Eq.~(\ref{eq:fYon}) is identical to the one obtained
in the usual Sullivan process with pseudoscalar meson--nucleon--hyperon
coupling \cite{Signal87, Holtmann96, Malheiro97}.

The $\delta$-function term $f_K^{(\delta)}$ arises from contributions
from kaons with zero light-cone momentum ($k^+ = 0$),
\begin{equation}
f_K^{(\delta)}(y)
= \frac{1}{\overline{M}^2}
  \int\!dk_\perp^2\,
  \log\Omega_K\, \delta(y)\,
  F^{(\delta)},
\label{eq:fKdel}
\end{equation}
where $\Omega_K = k_\perp^2 + m_K^2$, and $F^{(\delta)}$ is the
corresponding regulating function.
Note that the numerator in the on-shell function in Eq.~(\ref{eq:fYon})
depends on the hyperon mass $M_Y$ and not on the kaon mass, and hence
is labeled by the subscript $Y$.  In contrast, the integrand in the
$\delta$-function term is independent of the hyperon, and is labeled
only by $K$.

\subsection{Kaon bubble distribution}
\label{ssec:fyKbub}

Unlike the pseudoscalar theory, where only the rainbow diagram
appears, the pseudovector effective Lagrangian contains the
Weinberg-Tomazawa interaction, involving two kaon fields,
which give rise to the bubble diagram in Fig.~\ref{fig:loops}(b).
For a $K^+$ meson loop, the light-cone distribution associated
with the bubble graph is given by
\begin{eqnarray}
f^{\rm (bub)}_{K^+}(y)
&=& \frac{M}{f_\phi^2} \int\!\frac{d^4 k}{(2\pi)^4}\,
    \bar{u}(p) (-i\slashed{k}) u(p)
    \frac{i}{D_K}
    \frac{i}{D_K}
    2 k^+ \delta(k^+ - y p^+).
\end{eqnarray}
Performing the trace over the spinor indices, this can be written as
\begin{eqnarray}
f^{\rm (bub)}_{K^+}(y)
&=& \frac{i}{f_\phi^2} \int\!\frac{d^4 k}{(2\pi)^4}
    \frac{p \cdot k}{D^2_K}\, 2 y\,
    \delta\left(y-\frac{k^+}{p^+}\right).
\end{eqnarray}
Again using the identity in Eq.~(\ref{eq:identity}), the integrand
can be expressed in terms of a single kaon propagator, as for the
$\delta$-function term in Eq.~(\ref{eq:fKdel}),
\begin{equation}
f_{K^+}^{\rm (bub)}(y)
= 2 f_{K^0}^{\rm (bub)}(y)
= -\frac{\overline{M}^2}{(4\pi f_\phi)^2} f_K^{(\delta)}(y),
\label{eq:f-bub}
\end{equation}
where the relation between the $K^+$ and $K^0$ contributions
is made explicit.

\subsection{Hyperon rainbow distribution}
\label{ssec:fyH}

The coupling of the current to the hyperon in the rainbow diagram
in Fig.~\ref{fig:loops}(c) leads to the hyperon distribution
function given by
\begin{eqnarray}
\label{eq:fH-0}
f_{YK}^{(\rm rbw)}(y)
&=& M \frac{C^2_{KY}}{f_\phi^2}
    \int\!\frac{d^4 k}{(2\pi)^4}\,
    \bar{u}(p) (\slashed{k} \gamma_5)
    \frac{i(\slashed{p} - \slashed{k} + M_Y)}{D_Y}\,
    \gamma^+
    \frac{i(\slashed{p} - \slashed{k} + M_Y)}{D_Y}
    (\gamma_5 \slashed{k})\, u(p)			\nonumber\\
& & \hspace*{2cm} \times\ \frac{i}{D_K} \delta(k^+ - y p^+),
\end{eqnarray}
where one has two hyperon propagators and one kaon propagator.
Using the Dirac equation, Eq.~(\ref{eq:fH-0}) can be recast in the
reduced form
\begin{eqnarray}
\label{eq:fH-1}
f_{YK}^{(\rm rbw)}(y)
&=& - i \frac{C^2_{KY}}{f_\phi^2} \int\!\frac{d^4 k}{(2\pi)^4}
\left[
  \frac{\overline{M}^2 (k^2 - 2 y\, p \cdot k - 2 y M \Delta - \Delta^2)}
       {D_K D_Y^2}
- \frac{2 M \overline{M} y + 2 \overline{M} \Delta}{D_K D_Y}
- \frac{1}{D_K}
\right]						\nonumber\\
& & \hspace*{2cm} \times\
\delta\left(y - \frac{k^+}{p^+}\right).
\end{eqnarray}
The first term $(\sim 1/D_K D_Y^2)$ in Eq.~(\ref{eq:fH-1})
corresponds to the on-shell hyperon contribution, in analogy
with the on-shell term in the kaon rainbow contribution in
Eq.~(\ref{eq:fYon}).
The second term $(\sim 1/D_K D_Y)$ arises from the off-shell
components of the hyperon propagator, while the third term
$(\sim 1/D_K)$ involves the single kaon propagator and
contributes only at $k^+ = 0$.
It is convenient therefore to write the total hyperon rainbow
distribution function as a sum of three splitting functions
associated with the on-shell, off-shell and $\delta$-function
contributions,
\begin{equation}
f_{YK}^{(\rm rbw)}(y)
= \frac{C_{KY}^2 \overline{M}^2}{(4\pi f_\phi)^2}
  \left[ f_Y^{\rm (on)}(y) + f_Y^{(\rm off)}(y) - f_K^{(\delta)}(y)
  \right].
\label{eq:fYKrbw}
\end{equation}
The on-shell function $f_Y^{\rm (on)}$ is identical to that
in Eq.~(\ref{eq:fYon}), while the $\delta$-function term
$f_K^{(\delta)}$ is given by Eq.~(\ref{eq:fKdel}).
The additional off-shell splitting function in Eq.~(\ref{eq:fYKrbw})
is given by
\begin{equation}    
f_Y^{\rm (off)}(y)
= \frac{2}{\overline{M}} \int\!dk_\perp^2\,
  \frac{M y + \Delta}{(1-y) D_{KY}}
  F^{\rm (off)},
\label{eq:fYoff}
\end{equation}
where $F^{\rm (off)}$ is the corresponding off-shell regulating
function.  As with the on-shell function, the off-shell term also
contributes only at $y>0$, and depends only on the hyperon
(rather than kaon) mass.

\subsection{Tadpole distribution}
\label{ssec:fytad}

The distribution function associated with the tadpole diagram in
Fig.~\ref{fig:loops}(d), involving an operator insertion at the
$KKpp$ vertex, is given by
\begin{eqnarray}
f_{K^+}^{\rm (tad)}(y)
&=& -\frac{M}{f_\phi^2} \int\!\frac{d^4 k}{(2\pi)^4}\,
    \bar{u}(p) \gamma^+ u(p) \frac{i}{D_K} \delta(k^+ - y p^+),
\label{eq:f-tad-K+}
\end{eqnarray}
for the charged kaon loop, and
$f_{K^0}^{(\mathrm{tad})} = f_{K^+}^{(\mathrm{tad})}/2$
for the neutral kaon loop contribution.
Again using the Dirac equation, this can be written in terms
of the $f_K^{(\delta)}$ function as
\begin{equation}
f_{K^+}^{\rm (tad)}(y)
= 2 f_{K^0}^{\rm (tad)}(y)
= \frac{\overline{M}^2}{(4\pi f_\phi)^2} f_K^{(\delta)}(y),
\label{eq:ftad}
\end{equation}
so that the tadpole and bubble diagrams are in fact equal and
opposite \cite{Z1},
\begin{equation}
\label{eq:f-bub-f-tad}
f_K^{\rm (tad)}(y) + f_K^{\rm (bub)}(y) = 0.
\end{equation}

\newpage
\subsection{Kroll-Ruderman distribution}
\label{ssec:fyKR}

Because of the derivative coupling in the pseudovector theory,
by themselves the meson and baryon rainbow diagrams in
Figs.~\ref{fig:loops}(a) and (c) are not gauge invariant
(the sum of the bubble and tadpole diagrams, on the other hand,
is gauge invariant).
To ensure gauge invariance of all the chiral loop corrections
to the twist-two matrix elements requires, in addition, the
Kroll-Ruderman diagrams in Figs.~\ref{fig:loops}(e) and (f).
Inserting the relevant $pY$ operators in Eq.~(\ref{eq:Os}),
the light-cone momentum distribution associated with the
Kroll-Ruderman diagrams is given by
\begin{eqnarray}
\label{eq:f-KR}
f^{\rm (KR)}_{YK}(y)
&=& - i M \frac{C^2_{KY}}{f_\phi^2}
    \int\!\frac{d^4 k}{(2\pi)^4}\,
    \bar{u}(p)
    \left[ \slashed{k}\gamma_5
	   \frac{i(\slashed{p} - \slashed{k} + M_Y)}{D_Y}
	   \gamma^+ \gamma_5
	 + \gamma^+ \gamma_5
	   \frac{i(\slashed{p} - \slashed{k} + M_Y)}{D_Y}
	   \slashed{k} \gamma_5
    \right]
    u(p)				\nonumber\\
&& \hspace*{2cm} \times\ \frac{i}{D_K} \delta( k^+ - y p^+).
\end{eqnarray}
Applying the Dirac equation, the integrand can be decomposed
into two terms,
\begin{eqnarray}
f^{\rm (KR)}_{YK}(y)   
&=& - 2 i \overline{M} \frac{C^2_{KY}}{f_\phi^2}
    \int\!\frac{d^4 k}{(2\pi)^4}
    \left[ \frac{M y + \Delta}{D_K D_Y} + \frac{1}{M D_K}
    \right]
    \delta\left( y - \frac{k^+}{p^+} \right).
\end{eqnarray}
These can be identified with the off-shell and $\delta$-function
contributions from Eqs.~(\ref{eq:fYoff}) and (\ref{eq:fKdel}),
respectively, so that one has
\begin{eqnarray}
f_{YK}^{\rm (KR)}(y)
&=& \frac{C_{KY}^2 \overline{M}^2}{(4\pi f_\phi)^2}
    \left[ - f_Y^{\rm (off)}(y) + 2 f_K^{(\delta)}(y)
    \right].
\end{eqnarray}
Comparing the expressions for the kaon and hyperon rainbow
diagrams in Eqs.~(\ref{eq:fKYrbw}) and (\ref{eq:fYKrbw}),
one finds that the rainbow and KR splitting functions satisfy
the identity
\begin{equation}
f_{YK}^{(\rm rbw)} + f_{YK}^{(\rm KR)} = f_{KY}^{(\rm rbw)}.
\label{eq:fsum}
\end{equation}
Together with Eq.~(\ref{eq:f-bub-f-tad}), this guarantees that
the nucleon has zero net strangeness.  This will be evident when
we consider the convolution expressions for the strange and
antistrange PDFs in the nucleon in the next section.

\section{Strange PDFs in the nucleon: model-independent features}
\label{sec:model-indep}

Using the results for the nucleon $\to$ kaon $+$ hyperon splitting
functions in Sec.~\ref{sec:fy}, the generic convolution expression
in Eq.~(\ref{eq:conv}) can be written explicitly for the strange and
antistrange PDFs in the nucleon, incorporating the contributions
from all of the diagrams shown in Fig.~\ref{fig:loops}.
In this section we provide the formulas for the contributions to
the $s$ and $\bar s$ PDFs in terms of convolution of the rainbow,
Kroll-Ruderman, bubble and tadpole splitting functions and the
$s$ and $\bar s$ PDFs in the strange hadrons derived in
Sec.~\ref{sec:formalism}.
Following this we discuss the model-independent chiral nonanalytic
behavior of the moments of the $s$ and $\bar s$ PDFs, which is
required by the chiral symmetry of QCD.

\subsection{$s$ and $\bar s$ distributions}
\label{ssec:s-conv}

In the following we will assume for simplicity that the strange
and antistrange content of the nucleon arises exclusively through
the kaon loops in Fig.~\ref{fig:loops}, and that the bare nucleon is
made up entirely of nonstrange quarks.  In fact, strictly speaking
this constraint is not necessary for the discussion of the $s-\bar s$
asymmetry; the only requirement is that any non-chiral contributions
(perturbative or nonperbative) are symmetric with respect to $s$ and
$\bar s$.
The $\bar s$ PDF in the nucleon can then be written in terms of
convolutions of the kaon rainbow and kaon bubble splitting functions
from Figs.~\ref{fig:loops}(a) and (b), respectively, with the $\bar s$
distribution in the kaon \cite{XWang16},
\begin{eqnarray}
\bar{s}(x)
&=& \Big( \sum_{KY} f_{KY}^{(\rm rbw)}
        + \sum_{K}  f_{K}^{(\rm  bub)}
    \Big) \otimes \bar s_K,
\label{eq:sbar_conv}
\end{eqnarray}
where the rainbow terms are summed over
  $KY = K^+ \Lambda$, $K^+ \Sigma^0$ and $K^0 \Sigma^+$,
and the kaon bubble terms are summed over
  $K = K^+$ and $K^0$ for the proton initial state.

For the $s$-quark distribution in the nucleon, on the other hand,
the convolution involves the hyperon rainbow, kaon tadpole and
Kroll-Ruderman diagrams in Figs.~\ref{fig:loops}(c), (d) and
(e)--(f), respectively,
\begin{eqnarray}
s(x)
&=& \sum_{YK}
    \Big( \bar{f}_{YK}^{(\rm rbw)} \otimes s_Y
        + \bar{f}_{YK}^{(\rm KR)}  \otimes s_Y^{(\rm KR)}
    \Big)
 +\ \sum_K \bar{f}_K^{(\rm tad)} \otimes s_K^{(\rm tad)},     
\label{eq:s_conv}    
\end{eqnarray}
where the rainbow and Kroll-Ruderman contributions are again
summed over all $YK$ combinations, while the tadpole involves
a sum over $K^+$ and $K^0$.
For notational convenience, in Eq.~(\ref{eq:s_conv}) we define
the functions ${\bar f}_j(y) \equiv f_j(1-y)$.
This is necessary since we work in terms of the same momentum
fraction $y$ for all kaon and hyperon coupling diagrams in
Fig.~\ref{fig:loops}.
The strange quark hyperon PDFs, $s_Y$, are related to the $u$ and $d$
PDFs in the proton using SU(3) symmetry, as in Eqs.~(\ref{eq:sY}),
while the Kroll-Ruderman distributions, $s_Y^{(\rm KR)}$, are related
through SU(3) symmetry to the spin-dependent PDFs in the proton in
Eqs.~(\ref{eq:sKR}).  The tadpole distributions, $s_K^{(\rm tad)}$,
are given in Eqs.~(\ref{eq:s-tad}).
Note that with the convention of Eq.~(\ref{eq:Os}), the lowest
moments of all quark distribution functions in the hadronic states,
${\bar s}_K$, $s_Y$, $s_Y^{(\rm KR)}$ and $s_K^{(\rm tad)}$, are
normalized to unity.

\subsection{Leading nonanalytic behavior}
\label{ssec:LNA}

A defining feature of the chiral effective theory is the systematic
expansion of observables in power series in the meson mass, with
generally {\it a priori} undetermined coefficients.  However,
coefficients of terms in the expansion that are not analytic in
$m_K^2$ (such as odd powers of $m_K$ or logarithms of $m_K$) are
independent of the short-distance behavior of the theory and are
determined entirely by its infrared properties.
Any effective theory or model of QCD must therefore reproduce
exactly these coefficients, the most notable of which are the
leading nonanalytic (LNA) terms, if it is consistent with the
symmetries of QCD.
For moments of PDFs, the LNA terms were found previously
\cite{TMS00, Chen02, Arndt01} to have a characteristic
$m_\pi^2 \log m_\pi^2$ dependence (for pion loops), a feature
which was applied \cite{Detmold01} to analyze the chiral behavior
of lattice moments of the isovector quark PDFs.

In the present formulation, we can derive the LNA behavior of
the $n$-th moments of the individual $s$ and $\bar s$ PDFs,
defined as
\begin{subequations}
\label{eq:ssbarmom}
\begin{eqnarray}
S^{(n-1)}
&=& \int_0^1 dx\, x^{n-1}\, s(x),		\\
\overline{S}^{(n-1)}
&=& \int_0^1 dx\, x^{n-1}\, \bar s(x),
\end{eqnarray}
\end{subequations}%
and hence those of the $s-\bar s$ asymmetry, from the convolution
formulas (\ref{eq:sbar_conv}) and (\ref{eq:s_conv}) and the
nonanalytic properties of the splitting functions.
Of greatest phenomenological interest will be the $n=1$ and $n=2$
moments of the PDFs, which correspond to the number and momentum
sum rules.
The LNA behavior of the PDF moments is determined by the LNA
behavior of the moments of the splitting functions, each of
which can be expressed in terms of the three basic functions
$f_Y^{\rm (on)}$, $f_Y^{\rm (off)}$ and $f_K^{(\delta)}$ derived
in Sec.~\ref{sec:fy}.  We define the $n$-th moments of these,
integrated over the physical $y$ range, as
\begin{subequations}
\begin{eqnarray}
{\tilde f}_{{\rm on}, Y}^{(n)}
&=& \int_0^1 dy\, y^{n-1}\, f_Y^{(\rm on)}(y),		\\
{\tilde f}_{{\rm off}, Y}^{(n)}
&=& \int_0^1 dy\, y^{n-1}\, f_Y^{(\rm off)}(y),		\\
{\tilde f}_{\delta, K}^{(n)}
&=& \int_0^1 dy\, y^{n-1}\, f_K^{(\delta)}(y).
\end{eqnarray}
\end{subequations}
The LNA behavior is intrinsically infrared and is obtained by
considering the lower bound of the $k_\perp$ integration,
in each of the splitting functions.
Expanding in powers $m_K/M$ and $\Delta/M$, we find for the
$n=1$ moments,
\begin{subequations}
\label{eq:f1LNA}
\begin{eqnarray}
\overline{M}^2 {\tilde f}_{{\rm on}, Y}^{(1)}\Big|_{\rm LNA}
&=& (4 m_K^2 - 6 \Delta^2) \log m_K^2
   + 6 R\, \Delta\, \log\frac{\Delta-R}{\Delta+R},
\label{eq:f1onLNA}					\\
\overline{M}^2 {\tilde f}_{{\rm off}, Y}^{(1)}\Big|_{\rm LNA}
&=& -2 m_K^2 \log m_K^2
 -  \frac{2 R^3}{M_Y} \log\frac{\Delta-R}{\Delta+R},
\label{eq:f1offLNA}					\\
\overline{M}^2 {\tilde f}_{\delta, K}^{(1)}\Big|_{\rm LNA}
&=& -m_K^2 \log m_K^2,
\label{eq:f1delLNA}
\end{eqnarray}
\end{subequations}
where $R = \sqrt{\Delta^2 - m_K^2}$ and ${\cal O}(m_K/M, \Delta/M)$
corrections have been neglected.
For the $n=2$ moments of the splitting functions, we find the
LNA behavior
\begin{subequations}
\begin{eqnarray}
\overline{M}^2 {\tilde f}_{{\rm on}, Y}^{(2)}\Big|_{\rm LNA}
&=& \frac{4 \Delta}{3 M_Y} (-6 m_K^2 + 7 \Delta^2) \log m_K^2
 +  \frac{2 R}{3 M_Y} (5 m_K^2-14 \Delta^2)\,
    \log\frac{\Delta-R}{\Delta+R},
\label{eq:f2onLNA}					\\
\overline{M}^2 {\tilde f}_{{\rm off}, Y}^{(2)}\Big|_{\rm LNA}
&=& \frac{2 \Delta}{3 M_Y} (3 m_K^2 - 2 \Delta^2) \log m_K^2
 -  \frac{4 R^3}{3 M_Y}
    \log\frac{\Delta-R}{\Delta+R},
\label{eq:f2offLNA}					\\
\overline{M}^2 {\tilde f}_{\delta, K}^{(2)}\Big|_{\rm LNA}
&=& 0.
\label{eq:f2delLNA}
\end{eqnarray}
\label{eq:f2LNA}
\end{subequations}
Note that because the function $f_K^{(\delta)}(y) \propto \delta(y)$,
its $n=2$ and all higher moments vanish.

The LNA behavior of the $n$-th moments of the $\bar s$ PDF is
then given by
\begin{eqnarray}
\overline{S}^{(n-1)}_{\rm LNA}
&=& \frac{\overline{M}^2}{(4\pi f_\phi)^2}
\sum_{KY}
\Big[
  C_{KY}^2\, {\tilde f}_{{\rm on}, Y}^{(n)}
+ \left( C_{KY}^2 - 1 \right) {\tilde f}_{\delta, K}^{(n)}
\Big]_{\rm LNA}\,
\overline{S}^{(n-1)}_K,
\label{eq:sbarLNA}
\end{eqnarray}
where $\overline{S}^{(n-1)}_K$ are the moments of the $\bar s$ PDF
in the kaon, and the sums are taken over the appropriate hyperons
and kaons.

For the strange-quark PDF in the nucleon, because the convolutions
in Eq.~(\ref{eq:s_conv}) involve the splitting functions evaluated
at $(1-y)$, the expressions for the moments involve binomial sums
over the moments.  Specifically, one has
\begin{eqnarray}
S^{(n-1)}_{\rm LNA}
&=& \frac{\overline{M}^2}{(4\pi f_\phi)^2}
\sum_{KY}
\sum_{k=1}^n \binom{n-1}{k-1} (-1)^{k-1}	\nonumber\\
& & \hspace*{-2cm}
 \times
\left\{
  C_{KY}^2\,
  \Big[
    {\tilde f}_{{\rm on}, Y}^{(k)}
  + {\tilde f}_{{\rm off}, Y}^{(k)}
  - {\tilde f}_{\delta, K}^{(k)}
  \Big]_{\rm LNA}
  S^{(n-1)}_Y
 +\
  C_{KY}^2\,
  \Big[
    2 {\tilde f}_{\delta, K}^{(k)}
  - {\tilde f}_{{\rm off}, Y}^{(k)}
  \Big]_{\rm LNA}
  S^{(n-1)}_{\rm (KR) Y}
 -\
  \Big[
    {\tilde f}_{\delta, K}^{(k)}
  \Big]_{\rm LNA}
  S^{(n-1)}_{\rm (tad) Y}
\right\}.					\nonumber\\
& &
\label{eq:sLNA}
\end{eqnarray}
The expressions in Eqs.~(\ref{eq:sbarLNA}) and (\ref{eq:sLNA})
hold for any $n$, including $n=1$ and 2.
In particular, from Eqs.~(\ref{eq:an}), (\ref{eq:coeff_norm})
and (\ref{eq:barcoeff_norm}) each of the $n=1$ moments of
the PDFs in the strange hadrons is normalized to unity.
The LNA contributions to the $n=1$ moments for the strange and
antistrange distributions in the proton are therefore equivalent,
\begin{eqnarray}
S^{(0)}_{\rm LNA}\
&=& \frac{\overline{M}^2}{(4\pi f_\phi)^2}
\sum_{KY}
\Big[
  C_{KY}^2\, {\tilde f}_{{\rm on}, Y}^{(1)}
+ \left( C_{KY}^2 - 1 \right) {\tilde f}_{\delta, K}^{(1)}
\Big]_{\rm LNA}\
 =\ \overline{S}^{(0)}_{\rm LNA},
\end{eqnarray}
as required by strangeness conservation.
This is no longer the case for $n=2$ and higher moments, for which
the various moments in the strange hadrons $S^{(n-1)}_j$ are no
longer related.  Furthermore, for $n=2$ the antistrange quark moment
$\overline{S}^{(1)}$ depends on ${\tilde f}_j^{(2)}$, while the
strange quark moment $S^{(1)}$ depends on the combination
${\tilde f}_j^{(1)} - {\tilde f}_j^{(2)}$ from the combinatorics
in Eq.~(\ref{eq:sLNA}).

\section{Regularization}
\label{sec:reg}

For point particles, the functions $F^{\rm (on)}$, $F^{\rm (off)}$
and $F^{(\delta)}$ for the on-shell, off-shell and $\delta$-function
distributions in Eqs.~(\ref{eq:fYon}), (\ref{eq:fKdel}) and
(\ref{eq:fYoff}), respectively, are each set to unity, and the
integrations over the kaon loop momenta $k$ are ultraviolet divergent.
In the effective theory for the hadrons, which in nature always have
finite size, some regularization prescription must be adopted to render
the integrals finite.  In practice this is achieved by the regulating
functions aquiring momentum dependence such that the contributions
from large momenta are suppressed.

In the literature various prescriptions have been adopted, ranging
from dimensional regularization in traditional chiral perturbation
theory \cite{chiPT} to sharp cutoffs in $k_\perp$ \cite{Salamu15}
or form factors in more phenomenological approaches \cite{Speth98}.
Regularization with the help of a finite-range regulator has been
advocated \cite{Donoghue99, Wright00, Thomas03} as a practical method
which reflects the finite size of the baryon to which the chiral
field couples.
The effectiveness of the various prescriptions in providing accurate
results for expansions of various static properties of hadrons away
from the chiral regime have been discussed in Refs.~\cite{FRR1, FRR2}.

In any chosen regularization scheme it is important to respect the
symmetries of the underlying hadronic field theory, including Lorentz
invariance, gauge invariance, and chiral symmetry.  Schemes such as
dimensional regularization and PV regularization are
known to preserve both chiral and Lorentz symmetries, while for
other prescriptions some of these symmetries are not satisfied.
Simple application of hadronic form factors, for example, can lead
to problems with gauge invariance \cite{Z1, Faessler03}, and (in the
present application) with strangeness conservation in the nucleon.
Restoration of gauge invariance in the presence of form factors
requires the generalization of the chiral Lagrangian to include
nonlocal terms \cite{Faessler03, Terning91, Nonlocal16}.
Following the approach adopted in Ref.~\cite{XWang16}, here we
utilize the PV regularization method, which offers many of the
advantages of finite range regularization and preserves all of
the required symmetries.

\subsection{Pauli-Villars regularization}
\label{sec:reg-PV}

The PV regularization scheme involves subtracting from the point-like
amplitudes expressions in which the propagator mass is replaced by a
cutoff mass $\mu_1$, such that in the ultraviolet limit the differences
between the amplitudes vanish.
For the on-shell distribution $f_Y^{(\rm on)}(y)$, for example,
one replaces the $1/D_{KY}^2$ propagator in Eq.~(\ref{eq:fYon})
by $1/D_{KY}^2 - 1/D_{\mu_1}^2$, where $D_{\mu_1} = k^2 - \mu_1^2$.
This is equivalent to setting the regulating function $F^{\rm (on)}$
in Eq.~(\ref{eq:fYon})~to
\begin{equation}
F^{\rm (on)}
= 1 - \frac{D_{KY}^2}{D_{\mu_1}^2}.
\label{eq:Fon}
\end{equation}
Similarly for the off-shell hyperon function $f_Y^{(\rm off)}(y)$,
one replaces the propagator $1/D_{KY}$ in Eq.~(\ref{eq:fYoff}) by
$1/D_{KY} - 1/D_{\mu_1}$, in which case the off-shell regulating
function $F^{\rm (off)}$ is given by
\begin{equation}
F^{\rm (off)}
= 1 - \frac{D_{KY}}{D_{\mu_1}}.
\label{eq:Foff}
\end{equation}
For the $\delta$-function term $f_K^{(\delta)}(y)$, on the other hand,
because both the $k^-$ and $k_\perp^2$ integrations are individually
ultraviolet divergent, a single PV subtraction is not sufficient,
and two subtractions are necessary in the kaon propagator to render
the integrals finite,
\begin{equation}
\frac{1}{D_{K}}
\to \frac{1}{D_{K}}
  - \frac{a_1}{D_{\mu_1}}
  - \frac{a_2}{D_{\mu_2}}.
\end{equation}
Here the two subtraction constants $a_1$ and $a_2$ are related to the
cutoff masses $\mu_1$ and $\mu_2$ by
\begin{equation}
a_1 = \frac{\mu_2^2 - m_K^2}{\mu_2^2 - \mu_1^2},\ \ \ \ \
a_2 = \frac{\mu_1^2 - m_K^2}{\mu_1^2 - \mu_2^2},
\end{equation}
so that in the $k \to \infty$ limit the propagator term behaves as
$\sim 1/k^6$.  This leads to an effective regularizing function
in Eq.~(\ref{eq:fKdel}) given by
\begin{equation}
F^{(\delta)}
= 1 - \frac{a_1 \log\Omega_{\mu_1} + a_2 \log\Omega_{\mu_2}}
           {\log\Omega_K},
\label{eq:Fdel}
\end{equation}
with $\Omega_{\mu_i} = k_\perp^2 + \mu_i^2$.
The free parameters in the calculation are then the two cutoffs
$\mu_1$ and $\mu_2$, the constraints on which we discuss in the
following.

\subsection{Constraints on cutoff parameters}


Since the on-shell function, $f_Y^{(\rm on)}(y)$, depends only on
the $\mu_1$ cutoff parameter, the natural process to consider for
constraining $\mu_1$ phenomenologically is the inclusive production
of $\Lambda$ hyperons in $pp$ collisions.
For large values of the produced $\Lambda$ momentum ($1-y \gtrsim 0.7$)
and small $k_\perp \lesssim 100$~MeV, the dominant contribution to
the production process is expected to be from the exchange of a
single $K^+$ meson.
At larger kaon momenta $y$ (smaller $1-y$) multiple meson exchanges
and contributions from heavier meson and baryon intermediate states
will become more important \cite{Holtmann96, Kopeliovich12}.
These, however, cannot be computed within the chiral effective
theory framework and will not be considered here.

\begin{figure}[t]
\includegraphics[width=4.5in]{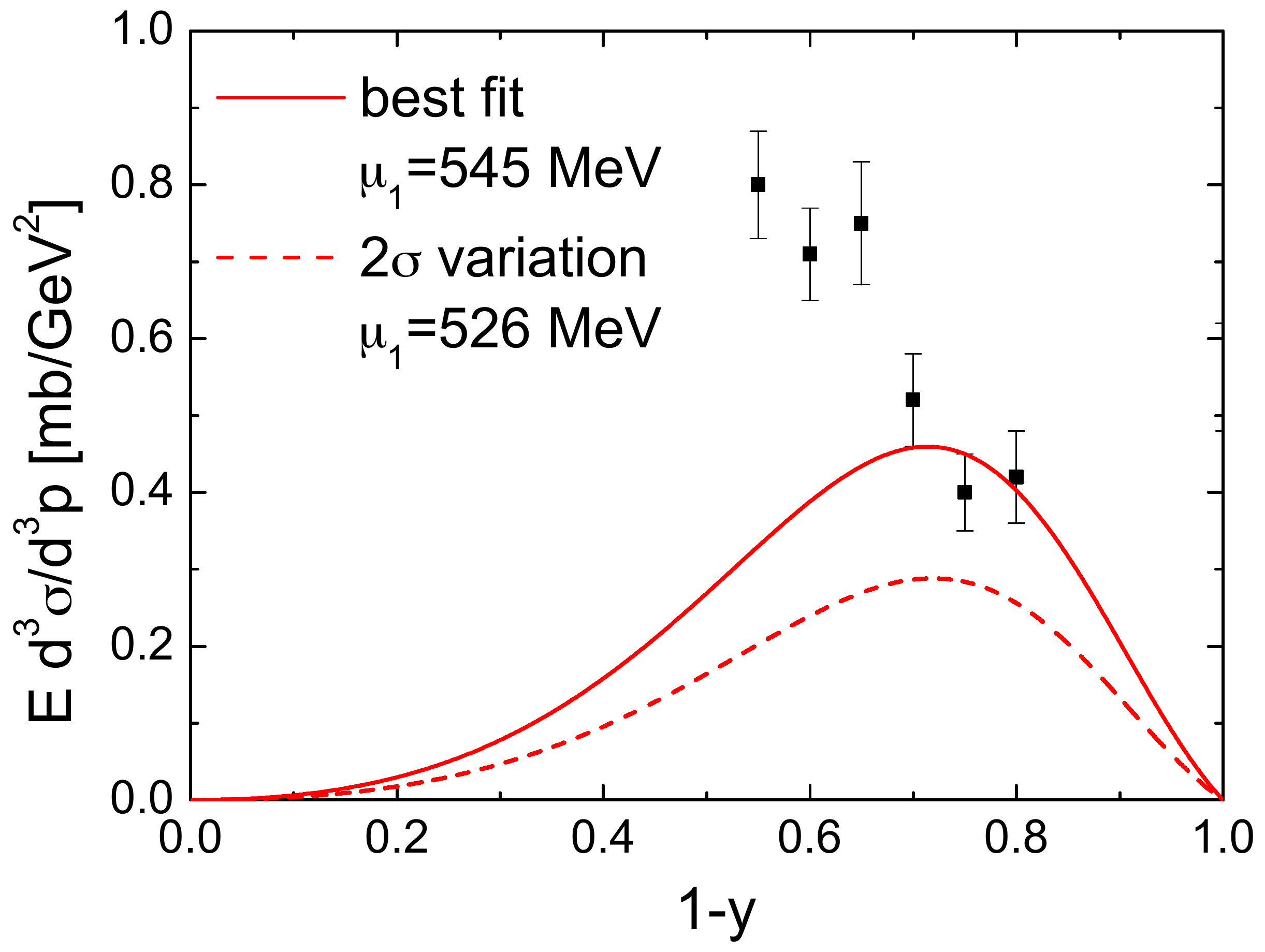}
\caption{Differential cross section for inclusive $\Lambda$ production
	in $pp$ scattering as a function of the momentum fraction $1-y$
	carried by the hyperon, for $k_{\perp}=75$~MeV~\cite{Blobel78}.
	The curves are fitted to the data at $1-y > 0.7$, with the
	best fit (solid line) obtained with the mass parameter
	$\mu_1 = 545$~MeV, and the fit $2\sigma$ below the central
	values (dashed line) with $\mu_1 = 526$~MeV.}
\label{fig:fit-Lambda}
\end{figure}

The differential cross section for the $pp \to \Lambda X$ reaction
with $K^+$ exchange is given by~\cite{Holtmann96}
\begin{eqnarray}
E \frac{d^3\sigma}{d^3p}
&=& \frac{C_{K^+ \Lambda}^2 \overline{M}^2}{16 \pi^3 f_\phi^2}
    \frac{y \left[ k_\perp^2 + (M y + \Delta)^2 \right]}
         {(1-y) D_{K^+ \Lambda}^2}
    F^{\rm (on)}(y,k_\perp^2)\, \sigma^{pK^+}_{\rm tot}(sy),
\label{eq:pp_sig}
\end{eqnarray}   
where $s$ is the $pp$ center of mass energy squared, and the total
$p K^+$ cross section $\sigma^{pK^+}_{\rm tot}$ is evaluated at
the $p K^+$ squared center of mass energy $sy$.
In Fig.~\ref{fig:fit-Lambda} the inclusive $\Lambda$ production
cross section data from Ref.~\cite{Blobel78} are shown as a function
of the hyperon momentum fraction $1-y$, for $k_\perp = 75$~MeV.
Taking the standard, constant value 
	$\sigma^{pK^+}_{\rm tot} = (19.9 \pm 0.1)$~mb~\cite{Povh92}
for the total $p K^+$ cross section in Eq.~(\ref{eq:pp_sig}),
we fit the $\mu_1$ parameter in the calculated cross section
to the data at small $y$ that are dominated by the lightest,
kaon-exchange contribution.
The best fit to data at $y < 0.3$ is obtained for the value
	$\mu_1 = (0.545 \pm 0.009)$~GeV,
where the error is statistical, giving a $\chi^2_{\rm dof} = 1.06$.
Extending the fitted range to $y < 0.4$ gives a significantly
worse fit, with $\chi^2_{\rm dof} \approx 3.7$, suggesting the
presence of other, non-kaonic contributions already for
$y \gtrsim 0.3$, consistent with the findings of previous
model-dependent analyses \cite{Holtmann96, Kopeliovich12}.
%
%
Including additional terms from non-kaonic backgrounds would in
practice reduce the magnitude of the kaon contributions allowed by
the data, so that the above cutoff can be taken as an upper limit.
As an estimate of the systematic uncertainty in this procedure,
we also consider a fit that lies two standard deviations below
the best fit, for which the cutoff parameter is $\mu_1 = 526$~MeV.

Additional constraints on the $\mu_1$ parameter can in principle
be obtained from comparisons of the $\bar s$ PDF in
Eq.~(\ref{eq:sbar_conv}) calculated from kaon loops with the
phenomenological $\bar s$ distribution extracted from global PDF fits.
The availability of antineutrino DIS data \cite{CCFR, NuTeV}, for
example, can isolate the $\bar s$ distribution from the $s$-quark
PDF, which contributes through the absorption of a $W^+$ boson
in neutrino DIS.  In practice, however, the uncertainties on the 
$\nu/\bar\nu$ data are typically considerably larger than those
on the corresponding electromagnetic cross sections.  Furthermore,
the neutrino measurements are usually performed on nuclear targets,
so that the cross sections must be corrected for nuclear effects,
which are not completely understood for neutrino scattering.
Thus, in practice little direct information exists on the $\bar s$
PDF from global analyses, which in fact usually assume symmetric
$s$ and $\bar s$ distributions.


On the other hand, the $s$-quark PDF is sensitive to the $\mu_2$ 
parameter in the $F^{(\delta)}$ function that regulates the
kaon tadpole contribution in Eq.~(\ref{eq:s_conv}).
Even though the splitting function associated with the tadpole
loop is a $\delta$-function at the kaon momentum fraction $y=0$,
Eq.~(\ref{eq:ftad}), the fact that the convolution (\ref{eq:s_conv})
involves a coupling at the hyperon vertex means that this contribution
to $s(x)$ in the nucleon will be proportional to $s_K^{(\rm tad)}(x)$.
Using the SU(3) relations in Eq.~(\ref{eq:s-tad}), this term will
then produce a valence-like shape that is nonzero at $x > 0$.
Comparisons with the phenomenological $s$-quark PDF as a function
of $x$ can then constrain the value of the $\mu_2$ parameter.

\begin{figure}[t]
\includegraphics[width=4.5in]{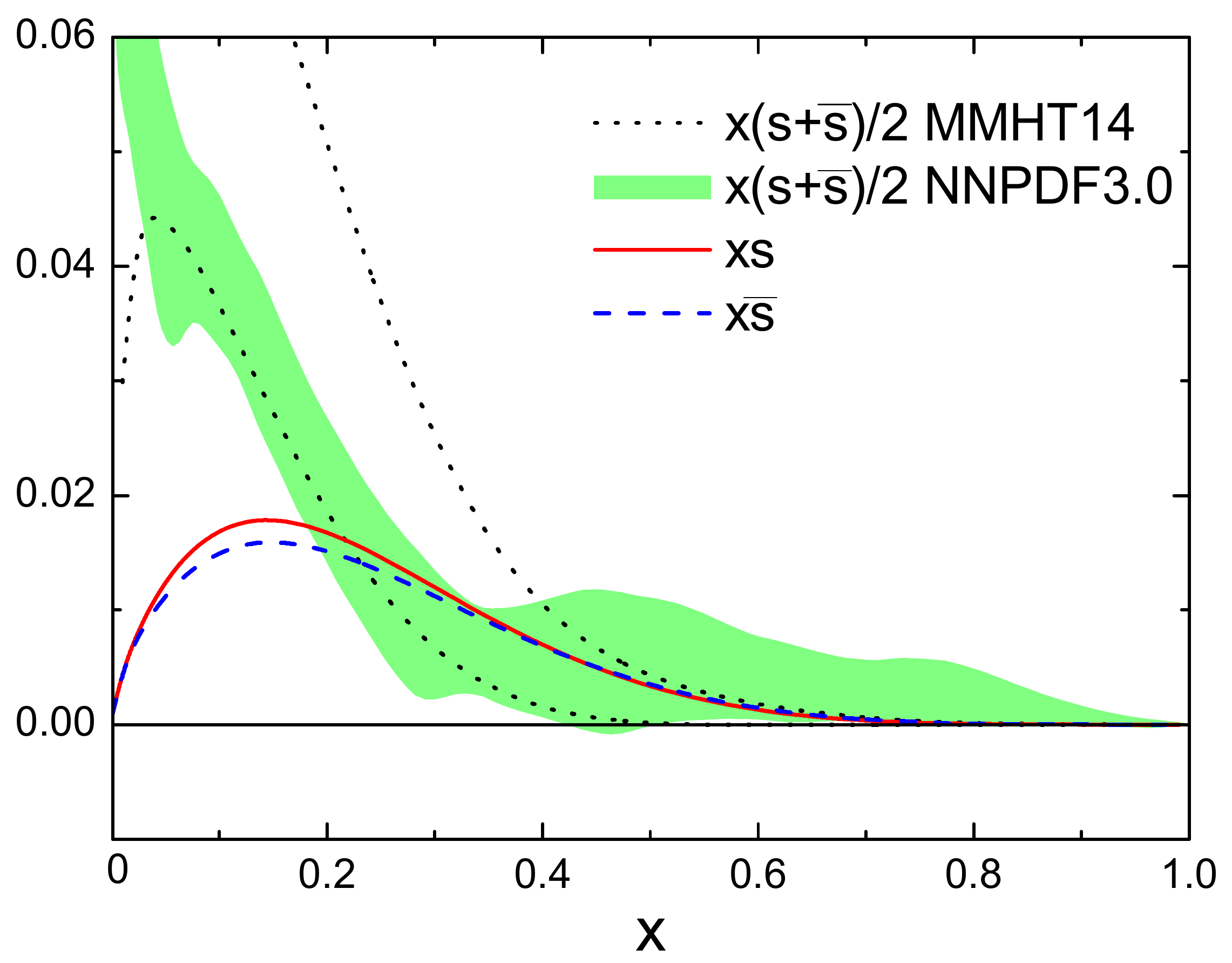}
\caption{Strange quark $xs$ (solid red curve) and
	antiquark $x\bar s$ (dashed blue curve) PDFs
	from kaon loops	for the best fit parameters
	$\mu_1 = 545$~MeV and $\mu_2 = 600$~MeV,
	compared with the phenomenological $x(s+\bar s)/2$
	distribution from the MMHT14~\cite{MMHT14}
	(black dotted curves) and NNPDF3.0~\cite{NNPDF3.0}
	(green shaded band) global analyses at $Q^2=1$~GeV$^2$.}
\label{fig:s+sbar}
\end{figure}

In Fig.~\ref{fig:s+sbar} the combined $s+\bar s$ distribution from
kaon loops is compared with several recent parametrizations from
global PDF analyses \cite{MMHT14, NNPDF3.0}.
In the evaluation of the $\bar s$ PDF in Eq.~(\ref{eq:sbar_conv}),
at the lowest order to which we work the strange quark PDF in the
kaon is related by SU(3) symmetry to the valence PDF in the pion,
	$\bar s_{K^+} = \bar s_{K^0} = \bar d_{\pi^+}$,
with the latter taken from a global PDF fit to $\pi N$ Drell-Yan
data by Aicher {\it et al.} \cite{Aicher10}.
For the strange quark PDFs in the hyperons, $s_Y$, and the strange
tadpole distributions, $s_K^{(\rm tad)}$, the SU(3) constraints in
Eqs.~(\ref{eq:sY}) and (\ref{eq:s-tad}), respectively, are used to
relate these to the $u$ and $d$ PDFs in the proton, for which the
parametrization by Martin {\it et al.} \cite{MRST98} is utilized.
For the strange KR distributions $s_Y^{(\rm KR)}$ at the $NKY$
vertex, on the other hand, Eqs.~(\ref{eq:sKR}) are used to express
these in terms of the spin-dependent PDFs in the nucleon, and in
practice we take the fit from Ref.~\cite{LSS10} for both the
polarized PDFs and the $D$ and $F$ values.
The results using other parametrizations for the spin-averaged
\cite{MMHT14, NNPDF3.0, CJ15} or spin-dependent \cite{DSSV09, JAM15}
$u$ and $d$ distributions yields very similar results.

The comparison of the $s$ and $\bar s$ PDFs in Fig.~\ref{fig:s+sbar}
calculated from kaon loops uses the maximum value of $\mu_1$
allowed by the $pp \to \Lambda X$ data in Fig.~\ref{fig:fit-Lambda},
and adjusts the maximum value of $\mu_2$ to ensure that the sum
$x(s+\bar s)$ does not exceed the phenomenological parametrization
at $Q^2=1$~GeV$^2$ within the quoted uncertainties,
	$(s + \bar s)_{\rm loops} \leq (s + \bar s)_{\rm tot}$.
Interestingly, while the MMHT14 parametrization \cite{MMHT14}
allows a slightly larger $s+\bar s$ at $x \lesssim 0.3$,
it places stronger constraints at larger $x$ values.
On the other hand, the NNPDF3.0 analysis, which uses a somewhat
different fitting methodology, gives slightly smaller strange
PDFs at low $x$, but permits a larger magnitude for $s+\bar s$
at $x \gtrsim 0.4$.
Taken as an approximately representative sample of the current
uncertainty on the strange quark PDF, the combined phenomenological
constraints in Fig.~\ref{fig:s+sbar} allow a maximum value for
the $\mu_2$ parameter of 600~MeV.
If we were to take the lower $\mu_1$ value from the inclusive
$\Lambda$ production data in Fig.~\ref{fig:fit-Lambda},
$\mu_1 = 526$~MeV, corresponding to the 2$\sigma$ deviation,
the loop contributions to $s+\bar s$ would remain consistent with
the phenomenological PDF constraints for $\mu_2$ as large as 894~MeV.

\newpage
\section{Strange asymmetry in the nucleon}
\label{sec:results}

Having obtained contraints on the $\mu_1$ and $\mu_2$ parameters
in our calculated $s$ and $\bar s$ PDFs from existing data on
inclusive $\Lambda$ production in $pp$ scattering and from
phenomenological PDFs, in this section we discuss in more detail
the implications of our results for the strange asymmetry in the
nucleon both as a function of $x$ and for the lowest moments.
We consider the two extremal cases for the cutoff parameters,
with the maximal $\mu_1$ from the $pp$ data combined with the
maximum $\mu_2$ from the comparison with the $s+\bar s$ PDFs,
	\mbox{$(\mu_1, \mu_2) = (545, 600)$~MeV},
and with a lower $\mu_1$ value for the 2$\sigma$ fit of the
$\Lambda$ production data and a correpondingly higher $\mu_2$ value,
	\mbox{$(\mu_1, \mu_2) = (526, 894)$~MeV}.
This range will provide a reasonable estimate of the systematic
uncertainty in our calculation.

\begin{figure}[t]
\includegraphics[width=4.5in]{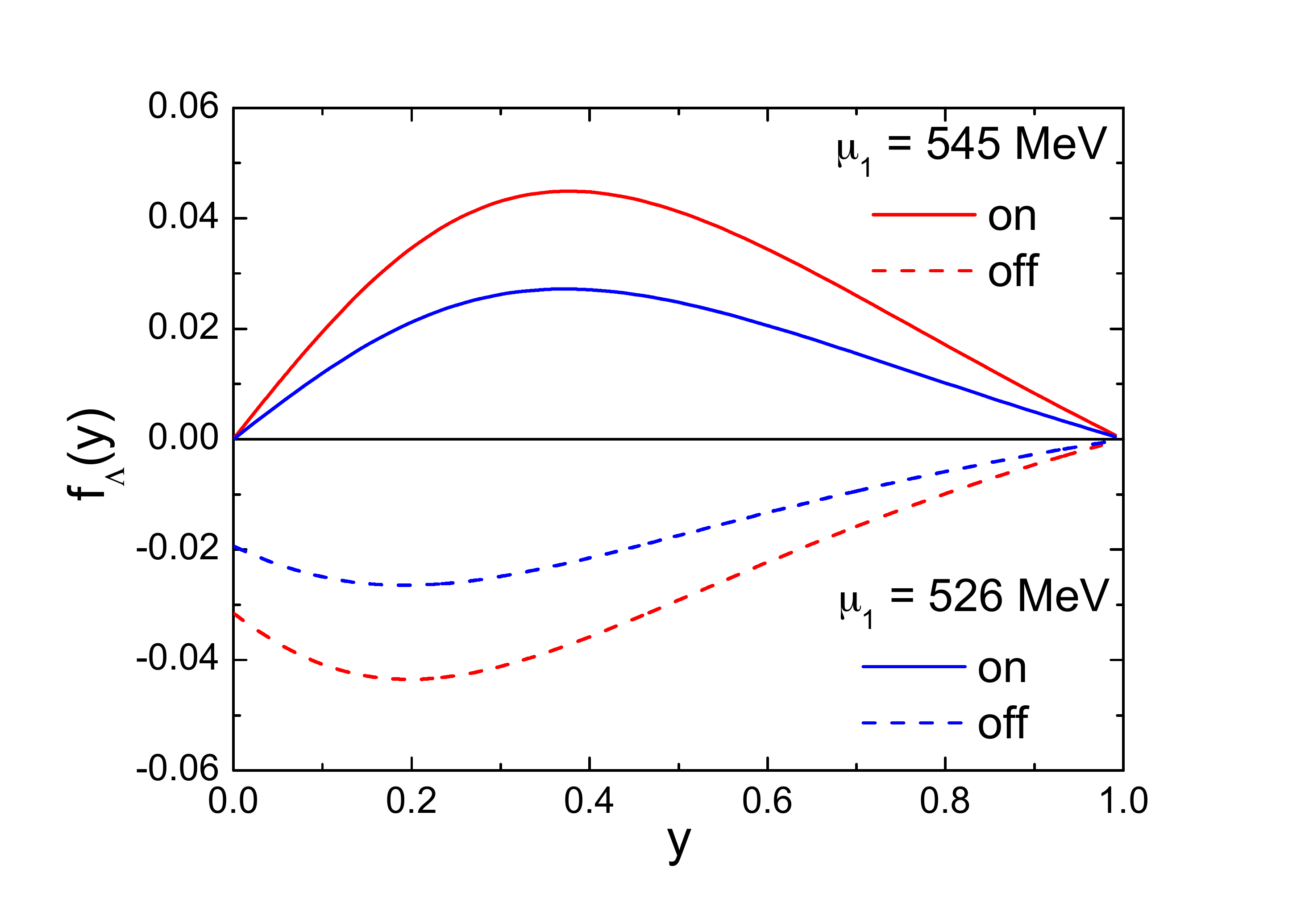}
\caption{On-shell (solid lines) and off-shell (dashed lines)
	contributions to the proton $\to K^+ \Lambda$ splitting
	functions for $\mu_1 = 545$~MeV (red curves) and
	$\mu_1 = 526$~MeV (blue curves).}
\label{fig:fy}
\end{figure}

To illustrate the variation for this range of cutoffs of the $KY$
splitting functions, in Fig.~\ref{fig:fy} we plot the on-shell and
off-shell functions $f_\Lambda^{\rm (on)}$ and $f_\Lambda^{\rm (off)}$
in Eqs.~(\ref{eq:fYon}) and (\ref{eq:fYoff}) for the
$p \to K^+ \Lambda$ dissociation as a function of $y$.
The on-shell distributions have a characteristic shape that peaks
around $y \approx 0.3-0.4$, with an obviously larger magnitude
for the higher cutoff, $\mu_1=545$~MeV.
Interestingly, the off-shell function is negative, with its
magnitude peaking at $y \approx 0.2$, and remains nonzero at $y=0$.
The latter result can be understood from the integrand of the
$f_Y^{\rm (off)}$ function in Eq.~(\ref{eq:fYoff}):
whereas for the on-shell function in Eq.~(\ref{eq:fYon}) the
$k_\perp$ dependence is multiplied by an overall factor $y$,
for the off-shell function the term in (\ref{eq:fYoff})
proportional to $\Delta$ remains finite in the $y \to 0$ limit.

\begin{figure}[t]
\includegraphics[width=4.5in]{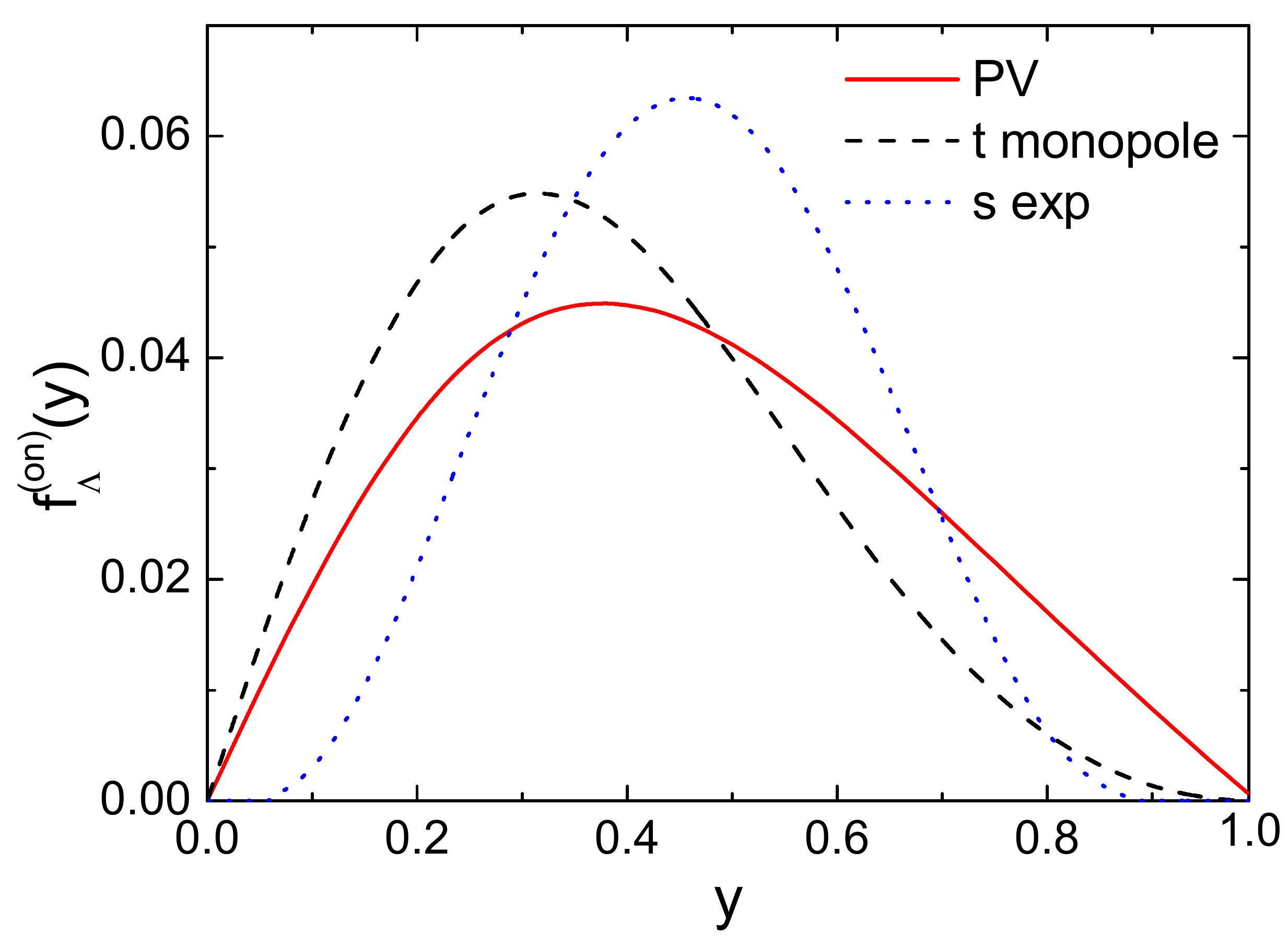}
\caption{Comparison of the on-shell proton $\to K^+ \Lambda$
	splitting function for the PV regulator with
	$\mu_1 = 545$~MeV (red solid curve) with the function
	computed with a $t$-dependent monopole form factor
	for $\Lambda_t = 0.928$~GeV (black dashed curve)
	and with an $s$-dependent form factor for
	$\Lambda_s = 1.293$~GeV (blue dotted curve), normalized
	to give the same value when integrated over $y$.}
\label{fig:fycomp}
\end{figure}

Note that the shape of the on-shell function in Fig.~(\ref{fig:fy}),
with the PV regulator, is qualitatively similar to the splitting
functions found in the literature which have been computed in terms
of form factors at the $NKY$ vertex \cite{Speth98}.  A comparison
of the $f_\Lambda^{\rm (on)}$ splitting functions computed with
PV regularization with the results obtained using $t$-dependent
\cite{Thomas83, Kumano91, MTS91, MST94, Kumano98} or $s$-dependent
\cite{Holtmann96, Malheiro97, Zoller92, MT93} form factors for the
function $F^{\rm (on)}$ is shown in Fig.~\ref{fig:fycomp}.
For the $t$-dependent form, the commonly used monopole shape is taken,
so that the function $F^{\rm (on)}$, which is the square of the form
factor, is a dipole,
\begin{eqnarray}
F^{\rm (on)}
&=& \left( \frac{\Lambda_t^2 - m_K^2}{\Lambda_t^2 - t} \right)^2,
\label{eq:Ftdep}
\end{eqnarray}
where
$t \equiv k^2
   = -[k_\perp^2 + y (M_Y^2 - (1-y) M^2)]/(1-y)$.
For the $s$-dependent form, an exponential shape is used,
\begin{eqnarray}
F^{\rm (on)}
&=& \exp\left(\frac{M^2 - s}{\Lambda_s^2}\right),
\label{eq:Fsdep}
\end{eqnarray}
where
$s \equiv (m_K^2 + k_\perp^2)/y
	+ (M_Y^2 + k_\perp^2)/(1-y)$.
The normaliation of each of the splitting functions is fixed to
be the same value as the PV-regulated form with cutoff mass
$\mu_1 = 0.545$~GeV, which is achieved with $t$-dependent
monopole cutoff mass parameter $\Lambda_t = 0.928$~GeV and
$s$-dependent exponential mass $\Lambda_s = 1.293$~GeV.

The shape with the PV regulator is slightly harder compared with
the other forms, but is closer to the $t$-dependent monopole at
low values of $y$.  Because of the $1/y$ and $1/(1-y)$ exponential
suppression in the $s$-dependent form factor, the result using
Eq.~(\ref{eq:Fsdep}) is significantly damped as $y \to 0$ and
$y \to 1$.

The $s$-dependent form in particular has been inspired in the
literature by attempts to satisfy $y \leftrightarrow 1-y$
symmetry relations between the splitting functions for the
kaon rainbow [Fig.~\ref{fig:loops}(a)] and
hyperon rainbow [Fig.~\ref{fig:loops}(c)] diagrams
\cite{Zoller92, Holtmann96}.
Namely, because of the kinematic relation 
	$s + t + u = M^2 + m_K^2 + M_Y^2$,
where $u \equiv (p-k)^2$, form factors that are functions of $s$
automatically satisfy the $t$- and $u$-channel crossing symmetry.
On the other hand, the $s$-dependent form is generally not Lorentz
invariant (it is invariant only under the light-cone longitudinal
and transverse boosts).
Furthermore, the use of momentum dependent form factors, whether
funtions of $t$ or $s$, is known to lead to a violation of gauge
invariance, requiring specific prescriptions to restore the
gauge symmetry through the introduction of nonlocal terms
\cite{Faessler03, Terning91, Nonlocal16}.
Calculations of PDFs using the splitting functions computed
with form factors on the basis of the local interactions in
Fig.~\ref{fig:loops}, let alone the rainbow diagrams by themselves,
are therefore in general not invariant under gauge or chiral
transformations.

\begin{figure}[t] 
\includegraphics[width=6.5in]{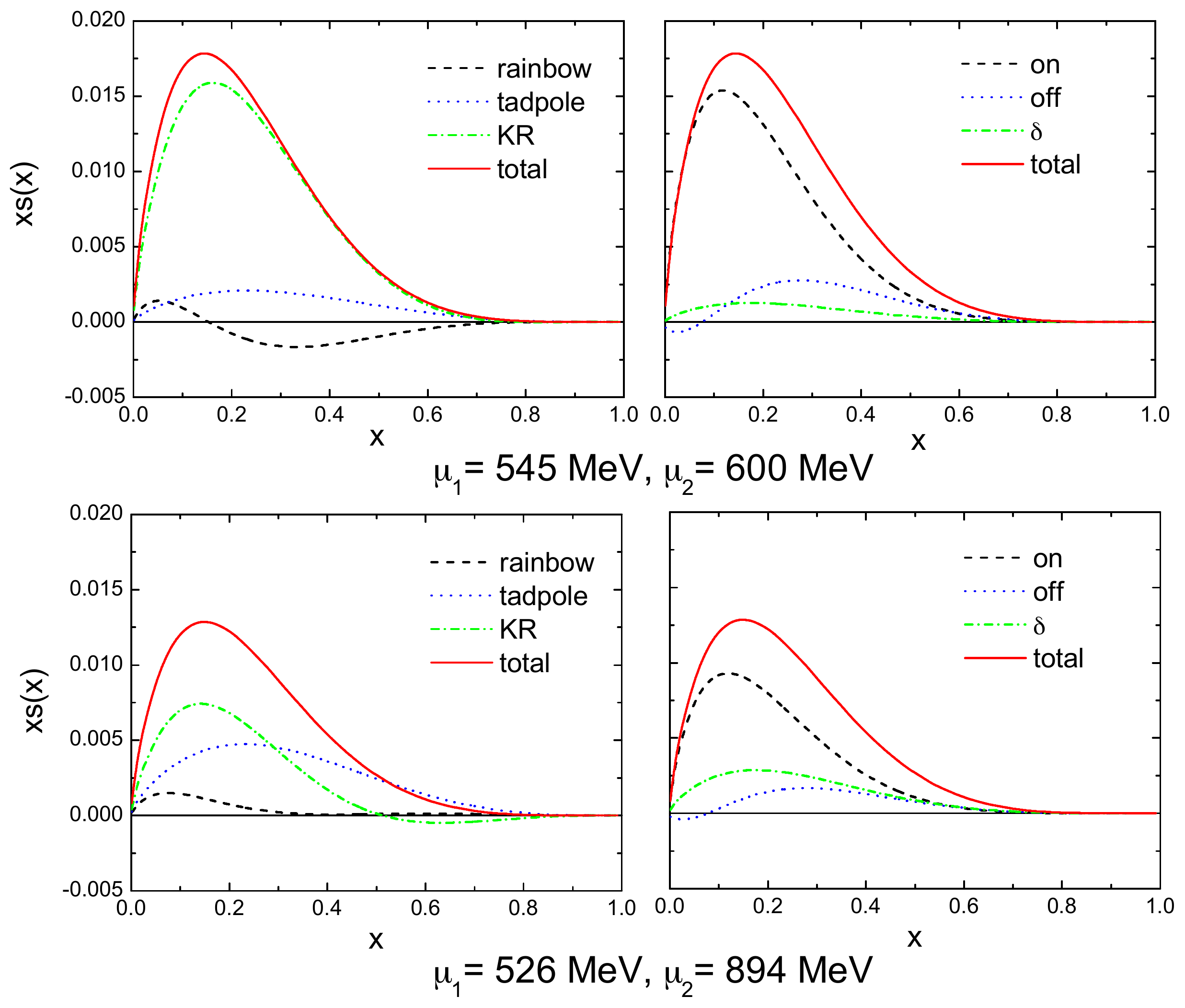}
\caption{Contributions to the $xs$ distribution at $Q^2=1$~GeV$^2$
	from various kaon loop diagrams in Fig.~\ref{fig:loops},
	for $(\mu_1, \mu_2) = (545, 600)$~MeV (top panels) and
	$(\mu_1, \mu_2) = (526, 894)$~MeV (bottom panels).
	The decomposition of the total into rainbow, tadpole and
	KR contributions (left panels) is contrasted with the
	decomposition into on-shell, off-shell and $\delta$-function
	contributions (right panels), according to
	Eqs.~(\ref{eq:s-comb})--(\ref{eq:sbar-comb}).}
\label{fig:case1-xs}
\end{figure}

It is instructive to quantify the relative contributions to the
strange-quark PDFs, as well as to their moments, arising from
the various diagrams in Fig.~\ref{fig:loops}.
As illustrated above in Fig.~\ref{fig:s+sbar}, the respective
magnitudes and shapes of the total contributions to $s$ and
$\bar s$ at $x > 0$ are similar, with $s$ slightly larger
than $\bar s$ at the peak around $x \approx 0.15$.
While only the on-shell piece contributes to $\bar s$ at $x > 0$
[Eq.~(\ref{eq:sbar_conv})], there are 3 contributions to the
$s$-quark PDF at nonzero $x$ [Eq.~(\ref{eq:s_conv})],
\begin{eqnarray}
s(x)
&=& \big( s^{\rm (on)} + s^{\rm (off)} + s^{(\delta)} \big)_{\rm rbw}\
 +\ s^{(\delta)}_{\rm tad}\
 +\ \big( s^{\rm (off)} + s^{(\delta)} \big)_{\rm KR}		\nonumber\\
&=& \underbrace{s^{\rm (on)}_{\rm rbw}}_{\rm on-shell}\
 +\ \underbrace{s^{\rm (off)}_{\rm rbw}
		+ s^{\rm (off)}_{\rm KR}}_{\rm off-shell}\
 +\ \underbrace{s^{(\delta)}_{\rm rbw}
		+ s^{(\delta)}_{\rm tad}
		+ s^{(\delta)}_{\rm KR}}_{\rm \delta-function}\, ,
\label{eq:s-comb}						\\
\bar{s}(x)
&=& \big( \bar{s}^{\rm (on)} + \bar{s}^{(\delta)} \big)_{\rm rbw}\
 +\ \bar{s}^{(\delta)}_{\rm bub}				\nonumber\\
&=& \underbrace{\bar{s}^{\rm (on)}_{\rm rbw}}_{\rm on-shell}
 +\ \underbrace{\bar{s}^{(\delta)}_{\rm rbw}
	      + \bar{s}^{(\delta)}_{\rm bub}}_{\rm \delta-function}\, ,
\label{eq:sbar-comb}
\end{eqnarray}
where we have suppressed the $x$ dependence in each of the terms
on the right-hand-side.
For the best fit parameters $(\mu_1, \mu_2) = (545, 600)$~MeV
(top panels in Fig.~\ref{fig:case1-xs}), the KR diagrams in
Figs.~\ref{fig:loops}(e)--(f) give the largest overall
contribution to $s(x)$, with the rainbow and tadpole
contributions relatively small.
Closer inspection of the various diagrams shows large cancellations
between the off-shell terms in the rainbow and KR diagrams, and
between the $\delta$-function terms arising from the rainbow, KR
and tadpole diagrams.
The net effect is that the total $s$-quark distribution is well
approximated by the on-shell part of the rainbow diagram, with the
total off-shell and $\delta$-function terms being relatively small.
This illustrates the vital role played by the tadpole and KR
diagrams, which are needed in a consistent theory along with the
rainbow contributions.
It also explains the phenomenological success of earlier
calculations of meson loop corrections to PDFs in terms of
on-shell rainbow contributions only.

For the alternative fit parameters from Sec.~\ref{sec:reg},
namely $(\mu_1, \mu_2) = (526, 894)$~MeV (bottom panels in
Fig.~\ref{fig:case1-xs}), the magnitude of the total
strange-quark PDF is slightly smaller, and the cancellations
between the various off-shell and $\delta$-function terms
are not as dramatic.  Nevertheless, even though the on-shell
part of the rainbow diagram does not saturate the total
contribution as completely, a similar qualitative behavior
is observed here also.

\begin{table*}[t]
\begin{center}
\caption{Individual contributions to the first ($n=1$) moments
	$S^{(0)}$ and $\overline{S}^{(0)}$ of the $s$ and $\bar{s}$
	PDFs from the diagrams in Fig.~\ref{fig:loops} at
	$Q^2=1$~GeV$^2$ for the two extreme cases considered,
	\mbox{$(\mu_1, \mu_2) = (545, 600)$~MeV}
	and \mbox{(526, 894)~MeV}.
	The moments are given in units of $10^{-2}$.\\}
\label{tab:1_moment}
\renewcommand{\arraystretch}{1.3}
\begin{tabular}{l|rr|rr}\hline\hline\
\ $(\mu_1,\mu_2)$\ \ 
  & \multicolumn{2}{|c|}{\ \ (545, 600)~MeV\ \ }
  & \multicolumn{2}{|c }{\ \ (526, 894)~MeV\ \ } 	\\
  & $S^{(0)}$
  & $\overline{S}^{(0)}$\ \
  & $S^{(0)}$  
  & $\overline{S}^{(0)}$\ \				\\ \hline
rbw (on)
  &   4.91  &   4.91\ \  &   2.97  &  2.97\ \		\\
rbw (off)
  & $-4.86$ &   ---\ \ \ & $-2.93$ &  ---\ \ \		\\
rbw ($\delta$)
  &   0.20  & $-0.20$\ \ &   0.47  & $-0.47$\ \		\\
\ &   \     &   \        &   \     &   \		\\
tad ($\delta$)
  &   0.59  &   ---\ \ \ &   1.36  & ---\ \ \		\\
bub ($\delta$)
  &  ---\ \ &  0.59\ \   &  ---\ \ &  1.36\ \		\\
\ &   \     &   \        &   \     &   \		\\
KR {\rm off)}
  &  4.86   &  ---\ \ \  &   2.93  &  ---\ \ \		\\
KR ($\delta$)
  &$-0.40$  &  ---\ \ \    & $-0.94$ &  ---\ \ \	\\ \hline
{\rm Total}
  &  5.30   &  5.30\ \   &   3.86  &  3.86\ \		\\ \hline\hline
\end{tabular}
\end{center}
\begin{center}
\caption{Contributions to the second ($n=2$) moments
	$S^{(1)}$ and $\overline{S}^{(1)}$ of the $s$ and $\bar{s}$
	PDFs from kaon loops at $Q^2=1$~GeV$^2$ for the two extreme
	cases considered, \mbox{$(\mu_1, \mu_2) = (545, 600)$~MeV}
	and \mbox{(526, 894)~MeV}.
        The moments are given in units of $10^{-3}$.\\}
\label{tab:x_moment}
\renewcommand{\arraystretch}{1.3}
\begin{tabular}{l|rr|rr}\hline\hline\
\ $(\mu_1,\mu_2)$\ \ 
  & \multicolumn{2}{|c|}{\ \ (545, 600)~MeV\ \ }
  & \multicolumn{2}{|c }{\ \ (526, 894)~MeV\ \ } 	\\
  & $S^{(1)}$
  & $\overline{S}^{(1)}$\ \
  & $S^{(1)}$  
  & $\overline{S}^{(1)}$\ \				\\ \hline
rbw (on)
  &   4.67  &   5.68\ \  &   2.83  &  3.41\ \		\\
rbw (off)
  & $-5.41$ &   ---\ \ \ & $-3.28$ &  ---\ \ \		\\
rbw ($\delta$)
  &   0.34  &  0\ \ \    &   0.79  &   0\ \ \		\\
\ &   \     &   \        &   \     &   \		\\
tad ($\delta$)
  &   0.95  &   ---\ \ \ &   2.21  & ---\ \ \		\\
bub ($\delta$)
  &  ---\ \ &   0\ \ \   &  ---\ \ &  0\ \ \		\\
\ &   \     &   \        &   \     &   \		\\
KR (off)
  &  6.35   &  ---\ \ \  &   3.85  &  ---\ \ \		\\
KR ($\delta$)
  &$-0.81$  &  ---\ \ \  & $-1.87$ &  ---\ \ \		\\ \hline
{\rm Total}
  &  6.10   &  5.68\ \   &   4.53  &  3.41\ \		\\ \hline\hline
\end{tabular}
\end{center}
\end{table*}

More quantitatively, the contributions of the various terms
to the moments of the $s$ and $\bar s$ PDFs are listed in
Tables~\ref{tab:1_moment} and \ref{tab:x_moment} for the
$S^{(0)}$, $\overline{S}^{(0)}$ and
$S^{(1)}$, $\overline{S}^{(1)}$ moments, respectively.
For the lowest ($n=1$) moments, the off-shell parts of the
rainbow and KR contributions to $S^{(0)}$ in fact cancel exactly,
leaving the on-shell component as the dominant term, and the
remaining contributions distributed among the $\delta$-function
pieces.  Strangeness conservation requires the on-shell
contribution to $\overline{S}^{(0)}$ to be identical to that for
$S^{(0)}$, with equivalent contributions from the tadpole and
bubble diagrams to the strange and antistrange moments, respectively.

\begin{figure}[t]
\includegraphics[width=6.5in]{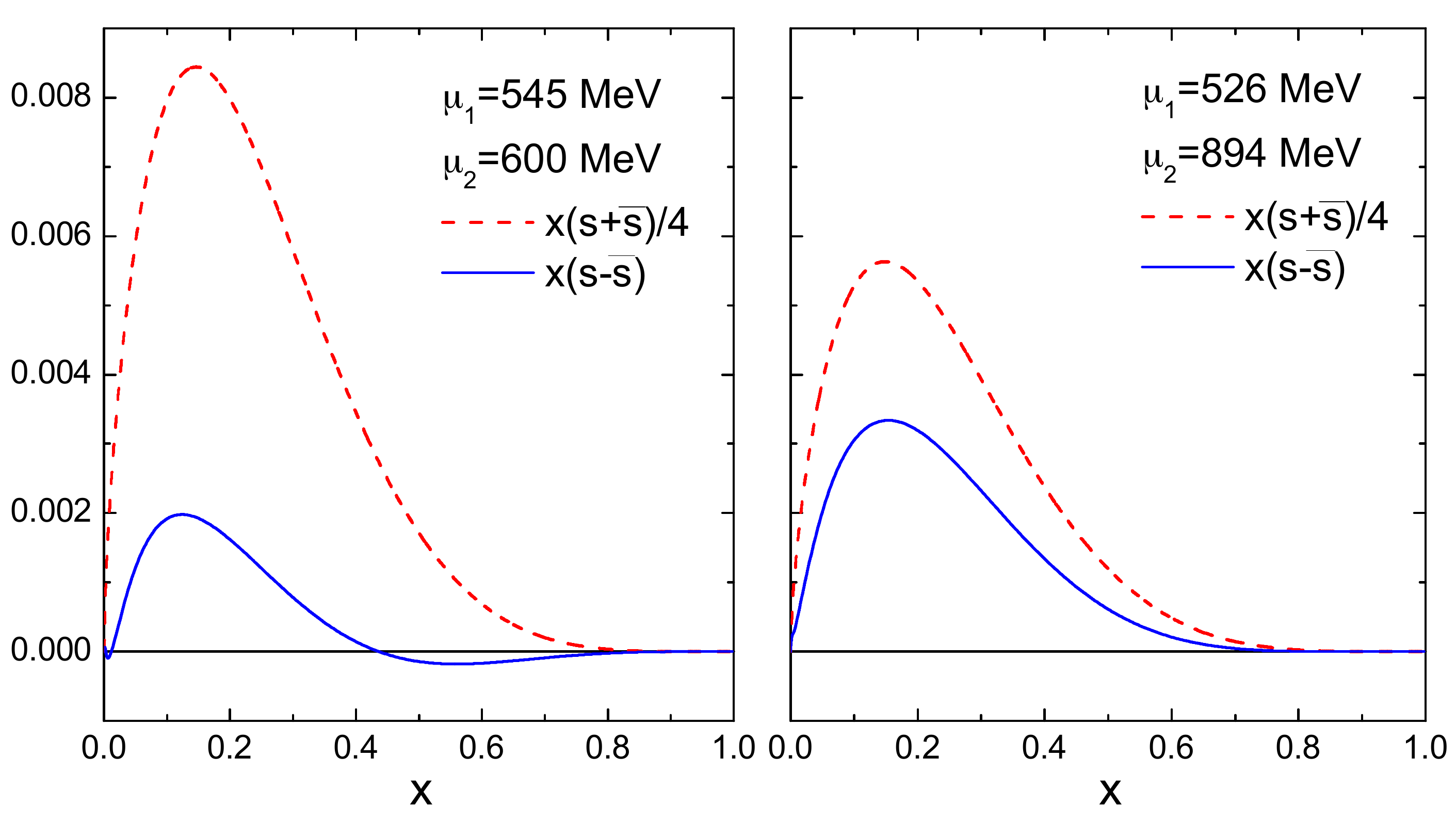}
\caption{Total sum $x(s+\bar s)$ (scaled down by a factor 1/4) and
	difference $x(s-\bar s)$ of the strange and antistrange PDFs
	from kaon loops at $Q^2 = 1$~GeV$^2$ with fit parameters
	$(\mu_1, \mu_2) = (545, 600)$~MeV (left panel) and
	$(\mu_1, \mu_2) = (526, 894)$~MeV (right panel).}
\label{fig:xssbar}
\end{figure}

For the second ($n=2$) moments in Table~\ref{tab:x_moment},
similarly large cancellations are observed between the off-shell
contributions to the $S^{(1)}$ moment from the rainbow and KR
diagrams.  Cancellations also occur between the positive
$\delta$-function parts of the rainbow and tadpole diagrams
with the negative $\delta$-function component of the KR diagrams.
In contrast, because of the additional power of $x$ in the $n=2$
moment definition, only the on-shell part of rainbow diagram
contributes to the $\bar s$ moment.
The net effect is thus a positive difference
$S^- \equiv S^{(1)} - \overline{S}^{(1)}$.
Note that while for the larger $\mu_1$ cutoff value both the
$S^{(1)}$ and $\overline{S}^{(1)}$ moments are bigger,
the difference
	$S^- = 0.42 \times 10^{-3}$
for $\mu_1=545$~MeV at $Q^2=1$~GeV$^2$ is smaller than for the
lower cutoff $\mu_1=526$~MeV, for which
	$S^- = 1.12 \times 10^{-3}$,
as is also apparent in Fig.~\ref{fig:xssbar}.
Here both the sum $x(s+\bar s)$ and difference $x(s-\bar s)$
are illustrated at $Q^2=1$~GeV$^2$ for both sets of cutoff values.
To display the sum and difference on the same plot, we scale the
much larger $x(s+\bar s)$ distribution by a factor 1/4.

For the best fit parameters
	$(\mu_1, \mu_2) = (545, 600)$~MeV,
the $x(s-\bar s)$ distribution peaks at around $x \approx 0.1$,
and has a zero crossing at $x \approx 0.45$, resulting in some
cancellation of the positive distribution at low $x$ and negative
distribution at large $x$.
Interestingly, for the
	$(\mu_1, \mu_2) = (526, 894)$~MeV
cutoff values, the asymmetry stays positive for all values of $x$,
with no zero crossing evident at $x>0$.  While this would not have
been possible in previous kaon loop calculations based on the on-shell
parts of the rainbow diagrams alone, Fig.~\ref{fig:loops}(a) and (c),
in the full chiral analysis strangeness is conserved through the
presence of the $\delta$-function contribution giving an overall
positive $\bar s$ at $x=0$, as evident in Table~\ref{tab:1_moment}.
This feature is not present in phenomenological PDF analyses of data,
which are sensitive only to the $x>0$ region.  Our observation of
nonzero $\bar s$ contributions increases the flexibility of data
analyses, by allowing a nonzero $s-\bar s$ distribution which
does not need to integrate to zero for $x>0$.

Note also that in Ref.~\cite{XWang16} the smallest difference
$S^-$ was found for the extreme case of $\mu_1=545$~MeV and the
minimal possible value of $\mu_2=m_K$.
For this value, the (generally positive) $\delta$-function
contribution to $s$ is rendered zero, thereby minimizing the
$s-\bar s$ difference.  While allowed phenomenologically, this
scenario appears less likely than the two cases considered above.

Finally, we can evaluate the effect of the $s-\bar s$ asymmetry
on the extraction of the weak mixing angle $\sin^2\theta_W$ from
the NuTeV data \cite{Zeller02}.  Folding the calculated PDFs
with the NuTeV acceptance functional, we find a correction
that lies in the range
$-7.7 \times 10^{-4} \leq \Delta(\sin^2\theta_W)
		     \leq
 -3.3 \times 10^{-4}$
at $Q^2=10$~GeV$^2$, corresponding to the range
$S^- = (0.42 - 1.12) \times 10^{-3}$ found here.
The negative $\Delta(\sin^2\theta_W)$ correction reduces the
overall discrepancy between the NuTeV value for the weak mixing
angle and the world average, but only by $\lesssim 0.5~\sigma$.
Our analysis therefore suggests that other explanations,
possibly involving an isospin dependent nuclear EMC effect
\cite{Cloet09} or charge symmetry violation in PDFs \cite{CSV},
may be more relevant in resolving the discrepancy \cite{Bentz10}.

\section{Conclusion}
\label{sec:conc}

Even after decades of study the quark--antiquark sea of the
nucleon offers both challenges and the potential for surprises.
The asymmetry between $\bar d$ and $\bar u$ antiquarks, with the
consequent violation of the Gottfried sum rule, is an obvious
example~\cite{NMC, E866}.
In this work we have focussed on the potential for an asymmetry
between the strange and antistrange quark PDFs in the nucleon.
Apart from relatively small effects arising at three-loop order
in perturbative QCD \cite{Catani04}, the dissociation of a nucleon
into a kaon and a hyperon, associated with the spontaneous breaking
of chiral SU(3) symmetry, is the natural source of such an asymmetry.

We have extended earlier studies of non-strange chiral corrections
to nucleon properties, in which the requirements of gauge invariance
and chiral symmetry were systematically explored.  Beyond leading
order in the chiral expansion this necessitates the inclusion of
Kroll-Ruderman terms, in addition to the usual rainbow diagrams
and tadpole contributions.
We have carefully explained the derivation of and presented formulas 
for the total contribution to the $s$ and $\bar s$ distributions at
next-to-leading order in the chiral expansion.
A novel feature of the calculation is the appearance of
$\delta$-function terms from kaon bubble diagrams, which contribute
to the $\bar s$ distribution at $x=0$.
These terms are independent of the ultraviolet regulator, and have
the important practical consequence that, in any experimental or
phenomenological study in which $x=0$ is inaccessible, the integral
of $s-\bar s$ will not vanish.

A further phenomenologically important consequence of the
$\delta$-function terms from the kaon tadpole diagram is that for
the $s$-quark distribution the corresponding splitting function
is a $\delta$-function at $\bar{y}=1$, where $\bar{y}$ is the
fraction of the nucleon momentum carried by the hyperon.
This leads to a valence-like component of the strange sea, which
cannot be generated from gluon radiation in perturbative QCD alone.

With the help of experimental data from inclusive $\Lambda$
production in $pp$ scattering and results from global PDF fits
\cite{MMHT14, NNPDF3.0}, we have obtained constraints on the
mass parameters for the Pauli-Villars regulators used in the
numerical calculation of the kaon loop contributions.
We find that $s$ and $\bar s$ quarks from this source contribute
up to $\sim 1\%$ of the total momentum of the nucleon, or
$\sim 30\%-50\%$ of the phenomenological strange sea of the
nucleon at a scale of $Q^2=1$~GeV$^2$ \cite{CJ15}.
In contrast, the magnitude of the strange asymmetry, $s-\bar s$,
is about a factor of 10 smaller than the sum.
Compared with other possible corrections to the NuTeV anomaly
\cite{Bentz10}, this is a relatively minor effect, reducing
the discrepancy by less than 0.5~$\sigma$.
The sign is, however, such as to reduce the anomaly,
which in itself answers a long-standing uncertainty.

Future improvements in the empirical determination of $s-\bar s$
could be obtained from higher precision deep-inelastic neutrino
and antineutrino scattering data from hydrogen or deuterium.
More immediately, perhaps, further constraints may be possible
through measurement of associated charm and weak boson production
in $pp$ scattering at the LHC \cite{Alekhin15}.
The theoretical framework utilized here can also be extended to
systematically explore the effects of kaon loops within the chiral
theory on strange quark polarization, including contributions from
both octet and decuplet hyperons, which will be discussed in a
separate publication \cite{Nonlocal16}.

\acknowledgments

We thank T.~J.~Hobbs and J.~T.~Londergan for helpful discussions
regarding many aspects of strange asymmetries in the nucleon,
and N.~Sato for helpful communications.
This work was supported by the DOE Contract No.~DE-AC05-06OR23177,
under which Jefferson Science Associates, LLC operates Jefferson Lab,
DOE Contract No.~DE-FG02-03ER41260, the Australian Research Council
through the ARC Centre of Excellence for Particle Physics at the
Terascale (CE110001104), an ARC Australian Laureate Fellowship
FL0992247 and DP151103101,
by CNPq (Brasil) 313800/2014-6 and 400826/2014-3,
and by NSFC under Grant No.~11475186, CRC 110 by DFG and NSFC.


\end{document}